# Nanoscale cuticle mass density variations influenced by pigmentation in butterfly wing scales

Deepan Balakrishnan[1,2†], Anupama Prakash[1†§], Benedikt J. Daurer[3†], Cédric Finet[1], Ying Chen Lim[2,4], Zhou Shen[2,4], Pierre Thibault[5], Antónia Monteiro[1*], and N. Duane Loh[1,2,4*]

[1]Department of Biological Sciences, National University of Singapore, Singapore. 117543.
[2]Centre for Bioimaging Sciences, National University of Singapore, Singapore 117557.
[3]Diamond Light Source, Harwell Science & Innovation Campus, Didcot OX11 0DE, UK.
[4]Department of Physics, National University of Singapore, Singapore 117551.
[5]Università degli Studi di Trieste, Trieste 34127, Italy.
[†]These authors contributed equally to this work.
[§]Current address: Centre for Engineering Biology and Department of Bioengineering, Imperial College London, SW7 2AZ, UK.
*Corresponding authors. Email: duaneloh@nus.edu.sg, antonia.monteiro@nus.edu.sg

## Abstract

How pigment distribution influences the cuticle density within a microscopic butterfly wing scale, and how both impact each scale's final reflected color, remains unknown. We use ptychographic X-ray computed tomography to quantitatively determine, at nanoscale resolutions, the three-dimensional mass density of scales with pigmentation differences. By comparing cuticle densities between two pairs of scales with pigmentation differences, we determine that the density of the lower lamina is inversely correlated with pigmentation. In the upper lamina structure of *Junonia orithya* and *Bicyclus anynana*, low pigment levels also correlate with sheet-like chitin structures as opposed to rod-like structures. Within each scale, we determine that the lower lamina in all scales has the highest density, and distinct layers within the lower lamina help explain reflected color. We hypothesize that pigments, in addition to absorbing specific wavelengths, can affect cuticle polymerization, density, and refractive index, thereby impacting reflected wavelengths that produce colors.

## Introduction

A detailed examination of the distribution of various matrix components within complex biological materials has often led to a better understanding of their color and other material properties [1–3]. Some of the main components found in arthropod cuticles are lipids, chitin fibres embedded in a matrix of cuticular proteins and tyrosine-derived pigment molecules, secreted by the underlying cells [4–7]. Variations in the chemical composition of these molecules and their organization within the layered matrix affect the properties and structure of the cuticle [6,8–14]. However, in the case of microscopic butterfly wing scales, which are the cuticular skeletons of single cells, little work has been done to explore how cuticle composition varies with scale morphology and color.

Arthropod cuticles are multi-layered, hierarchically organized structures [4,5]. Various layers are synthesized sequentially, with the outer epicuticular layers composed of lipids and proteins, while the inner procuticle layers are made of chitin and various proteins [15]. Chitin fibres are organized into diverse morphologies in cuticles of various natural systems [16,17] and their organization and interactions with other molecules can change within different parts of the same structure [16,18,19]. Chitin fibre organization can also change by modifications of the fibres such as deacetylation of chitin [20]. As in other cuticular structures, butterfly scale cuticles are similarly composed of multiple layers with different material compositions [21].

Butterfly scales owe their brilliant colors to either pigments in the cuticle [22,23] or the way that cuticle is patterned during development to produce non-pigmentary structural colors through light interference and diffraction [24,25]. As the scale grows, cuticle is secreted to the outside of the cell membrane to initially define two main cuticular surfaces: an upper lamina that is often intricately nanostructured, and a flat, sheet-like lower lamina (Fig 1A) [24,26]. Over time, the upper lamina is further sculpted into various sub-structures, forming longitudinal ridges first, followed by cross-ribs [27]. Lastly, the two laminas are connected by pillars or trabeculae that form towards the end of development [27,28]. Pigmentation only begins at the late stages of scale development [23,29].

Both upper and lower laminas of the scale can produce structural colors, but variations in just the lower lamina thickness are often sufficient to produce a large range of colors via thin-film interference [30]. It has recently become apparent, however, that optical models that predict structural colors from the physical dimensions of cuticle alone, assuming a homogeneous chitin matrix, are failing to capture the true color of the scales [30]. In addition, investigations into the genetics of butterfly scale development have identified a complex interplay between scale morphology and pigmentation [30–32]. Pigments, such as melanins, are often deposited in the cuticular mass of the scale, and knockouts of two melanin pathway genes in *Bicyclus anynana* butterflies led to decreased pigmentation levels in wing scales but also impacted scale structures [31]. In *yellow* crispants, an ectopic cuticular film was retained in the upper structures of black scales, closing the normally open 'windows' (Figs 1A, 1E), whereas in *Dopa Decarboxylase* (*DDC*) crispants, sheet-like trabeculae extended between the two laminas in place of pillar-like trabeculae [31]. Similar interactions between scale structure and coloration were observed in *Junonia* butterflies, where *Optix* crispant cover scales changed from brown to blue with a simultaneous decrease in pigmentation and an increase in the lower lamina thickness [30].

To better understand the influence of material composition on butterfly scale structure and their macroscopic optical properties, we require a 3D density map of the entire scale's structure at nanometer resolution with the corresponding color response that can be overlapped with the structure. However, there is no single imaging tool that can provide that. Various imaging modalities provide unique insights into the scale at different resolutions. Optical microscope images show us the color response, and a microspectrophotometer measures the reflectance and the absorbance spectra, but only with a few micrometers of resolution. Resin-section transmission electron microscopy (TEM) has a high sub-nanometer resolution that is adequate to resolve multilayers in the lower lamina, but due to heavy ion staining, the quantitative density estimation is not feasible. Optical modeling can combine the multilayers observed in the TEM images with the reflectance and absorbance measurements to suggest variations in material composition. Nevertheless, the absorbance measurement corresponds to total pigmentation and not pigment density.

Ptychographic X-ray computed tomography (PXCT) fills this gap by providing 3D mass density estimates through the collection of coherent diffraction patterns at overlapping scan positions across the specimen. The overlapping nature of the scans enables robust phase retrieval, which encodes fine structural details by recovering high spatial frequencies. Bright synchrotron sources, such as the Diamond Light Source, offer highly coherent and intense X-ray beams, enabling PXCT to achieve a few tens of nanometers resolution. Thus, PXCT can provide an intermediate resolution between the high-resolution TEM cross-sections and the low-resolution color response [33–37].

By utilizing a range of techniques from microspectrophotometry, scanning and transmission electron microscopy (SEM/TEM) to PXCT, we characterized how relative pigmentation differences are related to mass density distributions within different scale sub-structures, and their effects on scale color. In particular, we demonstrate the application of PXCT for estimating quantitative mass densities of scale sub-structures at nanoscale resolution. These measurements revealed density differences between the two laminas of the scale. In addition, the 3D structural measurements from PXCT enable the estimation of the absorption coefficient from the absorbance measurements that correspond to the pigment density. Further, distinct layers within the lower lamina suggest that cuticle heterogeneity within the scale might affect how light travels through or bounces off cuticular surfaces. A multi-layered optical model taking into account these compositional variations in the lower lamina predicts the experimental reflectance spectra of the scale more accurately than models based on a homogeneous cuticle matrix. Finally, the local tilt measurements of the lower lamina from PXCT correlating with the optical microscope images explain the spatial color variations in the scale and show the influence of the upper lamina structure in homogenizing the color.

## Results

### Pigment distribution is similar between the upper and lower laminas of all scale types

In this study, we compared two pairs of scales with differences in pigmentation in two species of nymphalid butterflies, *Junonia orithya* and *Bicyclus anynana*. In *J. orithya*, we compared the less pigmented blue cover scales of males with more pigmented brown female scales sampled from the same site (Figs 1B, 1D, S1A, S1B). Similarly, in *B. anynana*, we compared cover scales from the *yellow* mutant line with the more pigmented black cover scales of wildtype, sampled around the large eyespot (Figs 1C, 1E, S1C, S1D).

To determine the distribution of pigments between each scale's upper and lower laminas, we measured the reflectance and absorbance of single intact scales relative to scales with the upper laminas removed (Supplementary Data 1). Removing the upper lamina is beneficial, as the upper and lower laminas contribute differently to the total reflectance spectra. The lower lamina produces specular thin-film interference, and the upper lamina structures produce diffuse reflection (Lambertian reflection), so the optical responses can be inferred better by measuring the reflectance with and without the upper laminas. The upper laminas were scraped off with a pin, and the debris was removed using sticky tape (Fig 1F). The epi-illumination image shows that the color response of the lower lamina is not affected by the removal process, and the SEM image shows that there is no residue left (Fig 1F). The absorbance spectrum is affected by the scale's thickness and the density of color-absorbing pigments it contains. The thickness distributions of the upper and lower lamina from SEM and PXCT measurements (described below) enabled us to additionally estimate the absorption coefficient spectra (Figs 1D″, 1E″, S2), which is related to the pigment density (see Methods). Here, pigment density refers to the amount of pigment per unit volume, and this should not be confused with the cuticle mass density, i.e., the total mass of chitin and pigment per unit volume.

The reflectance of all scale types increased upon removal of the upper lamina (Figs 1D′, 1E′) while the absorbance decreased (Figs S1E, 1F). However, the absorption coefficient was similar between the upper and lower laminas (Figs 1D″, 1E″). This indicated that the density of light-absorbing pigments was similar between the two surfaces for each scale type.

**Pigmentation correlates with changes in scale morphology**

To assess scale structural changes associated with changes in pigmentation levels in all four scales, we measured inter-ridge and cross-rib spacing, and the thickness of the lower lamina, using a combination of SEM and FIB-SEM (Fig S3, Supplementary Data 2). The blue scales of *J. orithya* males differed from the highly pigmented brown scales of females by having greater inter-ridge (linear mixed effects model (lme), chisq = 336.12, p-value <0.001) and cross-rib distances (lme, chisq = 331.89, p-value <0.001), leading to larger 'windows' on the upper lamina (Figs S3A, S3B). In addition, the blue scales had a thicker lower lamina of ~190 nm as compared to the thinner lower lamina of ~158 nm in the brown scales (t-test, t(4) = 8.12, p <0.001) (Fig S3C). This thicker, thin film of males produced a structural blue via constructive interference [30].

Among the *B. anynana* pair of scales, the *yellow* mutant scales displayed an ectopic cuticular film on the upper lamina closing all the windows (Fig 1E). TEM cross-sections indicated that the film was a very thin cuticle layer of about 10-20 nm in thickness (Fig S4). These *yellow* mutant scales also had larger inter-ridge distances (lme, chisq = 12.91, p-value <0.001) and a denser arrangement of cross-ribs (lme, chisq = 10.82, p-value <0.001) relative to wildtype scales, as reported previously in (*21*) (Figs S3A, S3B). The thickness of the lower lamina of the *B. anynana* wildtype scales was similar to the *yellow* mutant scales based on FIB-SEM measurements (t-test, t(4) = 1.31, p-value = 0.23) (Fig S3C).

To probe the distribution of materials within the scale sub-structures at nanoscale resolution, we imaged one scale for all four scale types using ptychographic X-ray computed tomography (PXCT). A cryo-PXCT has been reported on a beetle scale to achieve a 3D resolution of 28 nm over a small sample volume of 7x7x7 $\mu m^3$ [37]. Here, we employed PXCT over much larger volumes of 70x35x35 $\mu m^3$ with a 3D resolution of 66.5 nm. The PXCT reconstructions provided 3D distributions of electron densities (Fig 2; see Methods, Supplementary Movies 1-4) (*26*). The 3D electron density maps showed both changes in density between upper and lower laminas as well as morphology between the pigmented and less pigmented scales of each species (Fig 2). In both *J. orithya* and *B. anynana* scales, the upper lamina of the less pigmented scale had sheet-like extensions from the cross-ribs as compared to the rod-like cross-ribs in the more pigmented scale (Figs 2, S5, S6). The *B. anynana yellow* mutant scale also had disordered and bent trabeculae (Figs S6B, S7), that are not seen in the wildtype scale (Figs S6A, S7). Decreased pigmentation correlating with sheet-like cross-ribs suggests that the concentration and the type of pigments may determine structural differences between the scales, i.e., rod-like or sheet-like cross-ribs and straight or bent trabeculae, by modulating chitin polymerization.

**Cuticle mass density is highest in the lower lamina within a scale, and the lower lamina of scales with less pigmentation have higher mass density**

To understand how the mass distribution varied between the different sub-structures of each scale, we computationally rotated and segmented the PXCT volumes into upper and lower laminas, and then the upper lamina was further clustered into the three sub-structures: cross-ribs, ridges, and trabeculae based on their column-wise (*z*-axis) 1D mass density profiles (Figs 3A-3D) (details in the step-by-step procedural description of mass density estimation section of Methods). Each such 1D profile shows the variation of mass densities along the abaxial-adaxial axis (top to bottom surface) of the scale, measured as voxels along the *z*-axis (Figs 3A-3D, inset). We segmented and classified these 1D profiles as lower lamina, ridges, cross-ribs, trabeculae, or ectopic film on the upper lamina (for *yellow* scales) and calculated a boxplot of average mass densities within each structural class for the four scale types (Figs 3A'-3D'). The mass densities within these sub-structures were calculated assuming each was made of only chitin since their cuticular protein, lipid and pigment compositions are unknown. Hence this would provide us a lower bound for the density estimates; a similar estimate based on melanin will increase the density values by ~11% (details in the segmentation and mass density estimation section of Methods). The contour-labels on the 2D density maps (scales seen from the top) validate the structural classification (Figs 3A''-3D''). The 1D mass density profiles shown in Figs 3A-3D represent the average 1D profiles of aligned and classified substructures from each scale type represented in Figs 3A''-D''. Since the classification and the density estimation are determined from 10,000 1D mass density profiles for each scale type, the statistical inference is more robust against noise. We caution that while PXCT reliably shows the relative cuticular densities between and within scales, the absolute cuticle density estimates reported here are lower bounds, owing to the well-known underestimation of densities in tomography of thin samples [38].

The limited resolution (66.5 nm voxel side-length) and artifacts in the PXCT reconstructions of the complex architecture of the *yellow* mutant scale resulted in ambiguous class boundaries between the ectopic film on the upper lamina, cross-ribs, and trabeculae. In Fig 3C, the inset arrows show the affected region in the 1D profiles that cause ambiguous class boundaries. Hence, the segmentation for this scale type was carried out based on the relative distance from the lower lamina. Further, we used double Gaussian fits to account for the presence of multiple substructures in a single 1D profile of the upper lamina (see Methods). The ambiguous regions were avoided in the density calculations and are represented as white regions in the density map (Figs 3C'' - 3D''). Moreover, because the short trabeculae under the cross-ribs were smaller than the voxel side-length of our PXCT reconstructions, these features were less apparent in both the *B. anynana* and *J. orithya* pairs of scales. Nevertheless, we determined the density from the trabeculae, ectopic film, and cross-ribs that are clustered unambiguously, as we have sufficient data points.

For the lower lamina, mass density was inversely correlated with pigmentation in both species. In all four scale types, the lower laminas were the densest sub-structures (Figs 3A-3D, 3A'-3D', 4), and scales with less pigmentation had higher lower lamina density.

For the upper lamina sub-structures, the relationship between density and pigmentation within each pair of scales was more complicated. In *B. anynana*, mass densities of the cross-ribs and trabeculae were higher in the lighter yellow mutant scale compared to the darker wildtype scale (Figs 3C', 3D'). However, the opposite was observed in the upper lamina of *J. orithya*, where cross-rib density was higher in the more pigmented female scale (Fig 3B'). The ridges of the more pigmented scales in each scale-pair had higher mass densities in both *B. anynana* and *J. orithya* (Fig 3).

PXCT's limited resolution impacted our mass density estimates. Recall that these densities were estimated from the Gaussian fits of the 1D *z*-profiles averaged within each sub-structure class identified using clustering of each scale. Since all the sub-structures (except the ectopic film, discussed below) are much larger in the *z*-axis (once rotated) than a single voxel (66.5 nm), this average density estimation along the *z*-axis is more robust against finite-volume effects than if we had estimated densities from single voxels (Fig S8). Conversely, this finite-volume effect impacted mass density estimates of the thin ectopic film on the yellow mutant *B. anynana*'s upper lamina. These films, which were identified in the regions between the sheet-like cross-ribs (yellow contours in Fig 3C''), were absent in the other three scale types in this study (Fig 4). Since these ectopic films were less than one voxel thick (red dashed box in Fig 3C'), they are more prone to finite-volume effects compared to other sub-structures, hence their mass density estimates were also more severely underestimated.

Overall, our PXCT reconstructions had sufficient resolution to compare the lower bound estimates of mass densities across the prominent sub-structures. Nevertheless, with modestly higher resolution (e.g., 10 nm), PXCT would likely be able to resolve the ectopic film and even the density variations within each sub-structure.

**Variation in cuticular composition within the lower lamina help to model blue structural color**

Whereas our PXCT reconstructions indicated variation in cuticlar mass densities amongst different sub-structures of the scale, further compositional variations were observed in the high-resolution TEM resin-sections (pixel size 0.8 nm) of the lower lamina of the *J. orithya* male scale alone (Fig 5A). Such multi-layered lower laminas have been observed in TEM sections of other butterfly species [39] and are likely related to the different cuticular layers such as epicuticle and procuticle made of different materials. Thayer and colleagues reported that the sub-200 nm thick lower lamina of blue butterfly scales should either be composed of a high refractive index material or require a more complex model to match the modeled and experimental reflectance spectra [30]. We confirmed that a single layer of chitin with refractive index 1.56 needs to be around 220 nm thick to match the observed blue reflectance spectra (Fig S9A), which is far thicker than our SEM and TEM measurements (Fig S3).

To understand how the compositional heterogeneity of the lower lamina impacts reflected color, we measured the reflectance spectra from regions of a blue scale from male *J. orithya* where we physically removed its upper lamina. The physical removal of the upper lamina leaves a planar lower lamina; the planar lower lamina of the male blue scales of *J. orithya* makes them a suitable candidate for modeling the reflectance spectrum (thin-film interference) to gain insights into the wavelength-dependent refractive index of the multiple-layers within, similar to Leertouwer and colleagues [40]. We fitted the measured spectra with a four-layer lower-lamina model, which matches the measured broad-spectral reflectance better than a homogenous chitin slab model, a homogenous melanin slab model, and a homogenous slab model with an intermediate refractive index (Fig 5B).

Our four-layer model accounted for the thickness variations in the lower lamina measured from SEM (Fig S3), the thickness of distinct layers within the lower lamina from the TEM measurements (Fig 5A), and the measured absorbance spectra (See Methods). Hence, only the scales with both the reflectance and the absorbance measurements are considered in the modeling (Fig 5B). Since the scales are placed on a glass slide in our measurements, we included scattering from the glass substrate in our models (Fig 5D).

Variations in the thickness of the lower lamina, the angular incident light caused by the numerical aperture, and backscattering from the background/substrate can cause incomplete destructive interference such that the spectrum's minimum does not fall to zero (See Methods, Fig S9). Specular reflection from wing scales that are improperly calibrated to a diffuse standard can result in overestimated reflectance. Hence, it is a common but perhaps ambiguous practice to subtract each measured spectra by an ad-hoc constant so that its resultant minima reaches zero, and divide this result by another ad-hoc correction factor to account for reflectance overestimation [30]. Since we included the glass substrate in the model, and instead used a mirror reference to calibrate the experimental spectra (see Methods, UV-VIS-NIR microspectrophotometry), we neither have to divide by a constant correction factor nor have to subtract for normalizing the spectra.

Our models empirically accounted for the changes in refractive index $n$ for different wavelengths of light $\lambda$ using Cauchy's transmission equation ($n(\lambda) = A + B/\lambda^2$). While our four-layer model regressed for different Cauchy coefficients ($A, B$) for each layer, these fitted coefficients were constrained to respective values between those of pure chitin and pure melanin. The four-layer model more accurately captures the measured reflectance spectra than a single-layer model and additionally enables inference of the spatial distribution of melanin-rich and chitin-rich regions within the lamina. The regressed Cauchy coefficients of each of the lower lamina layers from our model were: $A_{L1} = 1.677$, $B_{L1} = 2.68 \cdot 10^4$ nm$^2$, $A_{L2} = 1.529$, $B_{L2} = 1.52 \cdot 10^4$ nm$^2$, $A_{L3} = 1.588$, $B_{L3} = 8.84 \cdot 10^3$ nm$^2$, and $A_{L4} = 1.673$, $B_{L4} = 1.66 \cdot 10^4$ nm$^2$. The fitted refractive indices are higher for the top and bottom layers, corresponding to the darker TEM image layers (Fig 5A). From our model, the refractive index of L1 approximately matches that of melanin rodlets in bird feather barbules [41], while that of L2 is closest to chitin (Fig 5C).

We also noticed that spatial color variations were present within a single scale. To investigate which factors led to this, we first modeled overall scale color from the thin film interference reflectance and then subjected the same *J. orithya* male scale used in PXCT to high-resolution epi-illumination reflectance light microscope (Figs 5E, 5F). First, we applied our optical parameters found in the four-layer model fit discussed above to the lower lamina's spatially varying local tilt map extracted from the PXCT reconstruction. Here we assumed a constant thickness in local directions perpendicular to the lower lamina; hence the transverse variations of the lamina's apparent thickness as viewed from a fixed position either above or below the scale is entirely due to the lamina's changes in local tilt (i.e., undulations). We found that the simulated spatial variations in color reflectance across the scale's lower lamina (Fig 5G), without the upper lamina, were very sensitive to the changes in this local-tilt-induced apparent thickness (Supplementary Movie 5 (GIF)). The upper lamina structures act as a color diffuser that spatially homogenizes the variations in color reflectance when viewed from the abwing direction (Fig 5E). Overall, these results indicate that the lower lamina of the *J. orithya* blue scale is not a uniform thin film but an undulating, multi-layered structure composed of materials of different refractive indices that impact the resultant structural colors. The color is very sensitive to tilt and lamina thickness, but color is spatially homogenized by the light-diffusing sub-structures of the upper lamina [22,42].

## Discussion

In this work, we looked at two pairs of butterfly scales that differ in levels of pigmentation to understand how pigments affect cuticular densities and scale structure. By calculating the absorption coefficient, we show that though the absorbance decreases when the upper lamina of a scale is removed, the pigment per unit volume is similar between the upper and lower laminas. The amount of pigments affect cuticular structures, especially the cross-ribs, where sheet-like cross-ribs occur in scales with lower pigment levels as compared to rod-like cross-ribs in darker scales. We then identify that the lower laminas of all four scale types had the highest mass densities, and this inversely correlated with pigmentation.

The mass density variations between the scales in each pair of comparisons may be due to differences in their material composition or how their constituent monomers pack together. Due to the sequential nature of multi-layered cuticle development and the late deposition of pigments, distinct scale sub-structures that develop at different times may have combinations of the various cuticle layers with different amounts of pigments. For example, the last sub-structure to develop i.e, the trabeculae, likely lacks the epicuticle layer. Further, since pigments are deposited in scales during the late stages of development, the formation of the later sub-structures like the cross-ribs and trabeculae may be affected more by the presence of pigments.

The compositional heterogeneity of the lower lamina, seen in the TEM images and corroborated by our optical modeling, fits with the ordered deposition of various cuticle components. The darker regions in the TEM resin section (Figs 5A, S10), are caused by the osmium stain that we used as a post-fixative during sample preparation. Osmium preferentially binds to unsaturated, double-bonded molecules such as lipids and proteins, and pigments like melanins [43,44]. We considered the possibility that the observed contrast of distinctive layers within the lower lamina could result from incomplete penetration of osmium tetroxide, as staining prior to sectioning may lead to stronger deposition near the outer surfaces. If the penetration of osmium were limited to the outer surfaces, we would expect a symmetric diffusion from both sides of the lower lamina. However, we observe an asymmetrical distribution of contrast, which indicates that limited osmium penetration alone does not account for the observed contrast. (Figs 5A, S10). Additionally, the upper lamina structures exhibit strong and consistent staining throughout their thickness, suggesting that osmium tetroxide can penetrate the entire volume of the scale structure under our staining conditions (Fig S10). Though we cannot entirely rule out some staining artifacts, our observations are likely dominated by preferential osmium binding to melanin-rich substructures rather than the misleading contrast arising solely from limited stain penetration.
In our model (Fig 5D), the two darker layers (L1 and L4 in Fig 5D, corresponding to thicknesses *t1* and *t4* in Fig 5A) had higher refractive indices that are close to that of melanin's. We therefore speculate that the innermost layer L1 (closer to the trabeculae) contains melanins while the outer layer L4 is lipid- and protein-rich either with or without small amounts of melanins. The other two layers had lower refractive indices, closer to chitin, suggesting they might be the protein-rich (L3; Fig 5A) and the chitin-protein layers of the procuticle (L2; Fig 5A).

Our results suggest the following hypothesis. During early scale development, the outer layers of chitin-free epicuticle establish the ridges and the lower lamina. Chitin fibers are then produced during procuticle formation but due to the inherent differences in the shapes of the upper vs lower laminas (buckled ridges vs a flat lower lamina), packing and cross-linking of the fibers with proteins may be different between the two layers. Chitin may be more densely packed in the flat lower laminas leading to higher mass densities. Additionally, organization and cross-linking of the chitin fibers in the protein matrix potentially start earlier in the ridges and lower lamina compared to the cross-ribs and trabeculae. When pigments get incorporated at later stages of development, these pigment molecules have more spaces to infiltrate the more porous upper lamina sub-structures and may modulate the polymerization of chitin and proteins to a greater extent in the late upper lamina sub-structures (cross-ribs and trabeculae) compared to the lower lamina. Though the lower lamina receives the same amount of pigment molecules, due to the lack of porosity, the pigment molecules may not infiltrate the densely packed chitin matrix as much as in the upper lamina and may settle on top, forming the innermost layer of the lower lamina (L1; Fig 5A). Numerous examples of adsorbants and small molecules controlling nanocrystal growth have been established in the material sciences [45]. Pigment molecules like melanins might play similar roles in modulating chitin polymerization, and in differentiating rod-like and sheet-like structures.

Parameters other than pigmentation, however, may also regulate the variable cuticular density and morphology observed in the cross-ribs. Sheet-like cross-ribs developed in the less pigmented scales of both species (Fig S5, S6), but while the cross-ribs in the lighter *yellow B. anynana* mutant were denser than those in the darker wildtype scale (Fig 4C, 4D, S11), the opposite was seen for the lighter *J. orithya* male cross-ribs relative to those of the darker female scale (Fig 4A, 4B). It is possible that distinct cuticular proteins, or even different pigments, used by each species contribute to further regulate cuticle density and morphology of the cross-ribs.

In conclusion, by studying two pairs of differentially pigmented scales in two different species of butterflies, we analysed the broad distribution of pigmentation and its influence on scale structure and structural color. We identified that scales, just like other bio-composite materials such as wood and bone [46], exhibit variations in morphologies, structure, and composition across nanometer to micrometer ranges. Such hierarchical structuring provides multifunctionality at various structural levels by careful tuning of the material properties. We believe that the future to understanding the diversity of optical and mechanical properties of biomaterials will require 3D characterization of nanoscale variations in material composition, which quantitative density mapping techniques like PXCT, spectro-ptychography and Atomic Force Microscopy (AFM) complementing the traditional imaging modalities, can help uncover.

## Methods

### Specimens

Individuals of wildtype *Bicyclus anynana* butterflies were sampled from the stock population in the Monteiro lab. A mutant *yellow* line was generated by a CRISPR/Cas9 knockout of the gene *yellow* [47]. Individuals of *Junonia orithya* were collected from Coney Island, Singapore (Permit No: NP/RP 14-063-2) and were placed in mesh cages with the host plant, *Asystasia gangetica*, to collect eggs. They were fed with 5% sugar water and additional sources of heat were added to elicit activity. Caterpillars were reared for one generation in a 27°C temperature-controlled room with a 12:12 hour light cycle and 65% humidity.

### Optical imaging

Epi-illumination microscope images of individual scales (Fig 1) were obtained with a 20x lens of a uSight-2000-Ni microspectrophotometer (Technospex Pte. Ltd., Singapore) and a Touptek U3CMOS-05 camera. Images at different depths were obtained and computationally stacked using Adobe Photoshop v 22.5.1 (Adobe, California, USA). High-resolution epi-illumination images of the *J. orithya* male scale in Fig 5, were obtained with a 100x lens of a Nikon

ECLIPSE LV100ND microscope and a DS-Fi3 camera. Images at different depths were obtained and computationally stacked with the custom python code.

### UV-VIS-NIR microspectrophotometry

Reflectance and absorbance measurements from individual scales were captured with a 20x objective (NA = 0.5) from a ~2 μm sized spot using a uSight-2000-Ni microspectrophotometer (Technospex Pte. Ltd., Singapore). Illumination was from a Mercury-Xenon lamp (ThorLabs Inc., New Jersey, USA). Due to the glass microscope objective, the microspectrophotometer spectra were unreliable for wavelengths in the UV range (< 350 nm). The reflectance measurements were taken from five different scales of *J. orithya* male (blue) and four different scales of female (brown), two scales of *B. anynana* wildtype, and a single scale of *B. anynana yellow* mutant. The absorbance measurements were taken from two different scales of *J. orithya* male (blue) and a single scale of *J. orithya* female (brown), *B. anynana* wildtype, and *yellow* mutant. All the scales are collected from a single individual of each type. Since we plotted the average reflectance curves from multiple scales with shifted peaks in the reflectance data, the reflectance curves are dimmer and bi-modal in Fig 1D'-1E'.

To remove the upper lamina, the scales were placed on a glass slide, and another glass slide was placed on top to partly cover each scale. The upper lamina in the exposed regions of the scales was scraped off with a pin, and the debris was removed using sticky tape. An SEM image and an optical image of a scale with regions of intact and cleanly peeled off upper lamina are shown in Fig 1F. For the reflectance measurements, we used a mirror as a reference (made of polished aluminium, manufactured in the Singapore University of Technology and Design) because the lower lamina of scales are specular reflectors. As the mirror reference reflects all the incident light, it can model reflectance quantitatively.

For the absorbance measurements, a drop of refractive index matching fluid (clove oil, Hayashi Pure Chemical Ind., Ltd.) was added on top of the lower lamina removed scales and covered with a cover slip (Fig S1). The region under the cover slip with no scale was used as a reference for absorbance measurements. Analysis and spectral plots were done in R Studio 1.4.1106 with R 4.0.4 [48] using the R-package *pavo* (v 2.7) [49].

### Scanning electron microscopy (SEM) and scale measurements

Three to six individual scales were sampled from the different areas under study from five different individuals and placed on carbon tape. Samples were processed in one of two ways: a) sputter coated with gold using a JFC-1100 Fine Coat Ion Sputter (JEOL Ltd. Japan) and imaged using a JEOL JSM-6510LV scanning electron microscope (JEOL Ltd. Japan) or b) sputter coated with platinum using a JFC-1600 Auto Fine Coater (JEOL Ltd. Japan) and imaged using a JEOL JSM-6701F FESEM (JEOL Ltd. Japan).

All measurements were done using ImageJ 1.53p [50]. For inter-ridge distance measurements, five individual measures were taken for each scale and averaged. Cross-rib distances were calculated by measuring the number of cross-ribs per 15 μm.

Statistical analysis

Analyses were done on R Studio 1.4.1106 with R 4.0.4 [48]. Since the datasets were hierarchical with many scales sampled from each individual, the differences in mean values of inter-ridge and cross-rib distances among the different comparisons were modeled using a linear mixed-effects model (lme). This type of model allows for fixed and random effects where we considered Sex (*J. orithya* male or female) and Type (*B. anynana* WT or *yellow* mutant) as fixed factors and scale nested within individual as a random factor. The linear mixed-effects model was run using the nlme package (v 3.1.152) [51]. To address the violation of the homogeneity of variance in certain measurements, we used the 'varIdent' function in the nlme package to allow for different variance structures. Models were compared using the ANOVA function, and the best models were selected using the Akaike Information Criterion.

Focused ion beam scanning electron microscopy (FIB-SEM) and thickness measurements

Lower lamina thickness of all four scale types was measured using FIB-SEM. Measurements were corrected for tilt by dividing the values by sin 52 °. Samples were first sputter-coated with platinum, then milled at the centre of each scale using a gallium ion beam on a FEI Versa 3D. The following settings were used: beam voltage −8kV, beam current −12pA at a 52° tilt. Images were acquired with the following settings: beam voltage −5kV, beam current −13pA. Five scales were sampled from a single individual of each category. For each scale, ten measurements of thickness were taken along the cross-section of the lower lamina using Fiji 1.53p [50].

Ptychographic X-ray computed tomography (PXCT)

In preparation for the tomographic X-ray imaging, one scale of each scale type of both *J. orithya* and *B. anynana* were fixed at the tip of a sharp needle (0.4 mm diameter at the base), which was then inserted into a 0.5 mm hole drilled at the center of a standard SEM specimen stub (12.5 mm diameter, 3.2 x 8mm pin). Additionally, the needle was glued to the stub at its base. The PXCT data was collected at the I13-1 Coherent Branchline [52–54] at the Diamond Light Source with a photon energy of 9.7 keV, similar to a previous experiment on butterfly scales [35]. The sample plane was located ~14 mm downstream of the focal plane of the X-ray beam providing a defocused beam with a diameter of ~13 um. Focusing was achieved using a Blazed Fresnel zone plate (courtesy of collaboration with C. David/PSI) with 400 um diameter and 150 nm outer zone width. The data for the male specimen of *J. orithya* was collected with the Merlin detector, all other specimens were collected with the EXCALIBUR detector [55,56]. Both detectors are based on Medipix3 arrays with 55 um pixel pitch, and they were both located 14.65 m downstream of the sample plane. For each PXCT measurement, the specimen stub was mounted at the center of the translational and rotational sample stage and scanned transversally with a step size of 4 um for 1250 equally spaced tomographic angles between 0° and 180° providing a ptychographic dataset for each tomographic angle. The ptychographic data was reconstructed using 1000 iterations of the semi-implicit Douglas-Rachford algorithm [57] with sigma = 0.5 and tau = 0.1 as implemented in the PtyPy framework [58]. As a pre-processing step, all diffraction patterns were cropped to 512x512 pixels resulting in a reconstructed pixel size of 66.5 nm in the given experimental geometry, and a bad pixel mask (including hot, dead, inactive pixels) was used. The initial guess for the illumination was determined by collecting and reconstructing data from a Siemens star test target. For each specimen a tomographic volume was reconstructed based on the ptychographic phase projections using the filtered-back projection algorithm (Ram-Lak filter) from the ASTRA toolbox [59] as implemented in the Savu framework [60]. Prior to the tomographic reconstruction, the phase projections were aligned along the vertical direction by correlating integrated vertical line profiles [61], and the center of rotation was determined using a simplified version of the method described [62] but with a manual search instead of the Fourier analysis. To improve the quality of the tomographic reconstructions, the volume was forward-projected to generate a set of synthetic phase projections. These synthetic phase images were compared against the original phase reconstructions while searching for an improved alignment in both the vertical and horizontal directions [61,63]. The updated phase projections were subsequently put through the same tomographic reconstruction process as described above. This iterative alignment scheme was repeated 5 times providing a total of 4 three-dimensional electron density maps (Fig 3).

Segmentation and mass density estimation

We used a series of computational techniques to process the reconstructed volumes from PXCT to extract quantitative data effectively. We illustrate the density estimation procedure with a detailed step-by-step schematic in Fig 6, using a small cross-section of the PXCT reconstructed volume of *J. orithya* blue male scale (Fig 6A). The reconstructed volumes were first rotated such that the lower lamina aligns flat (as much as possible) on the *x-y* plane (Fig 6B). This rotation was performed using a combination of principal component analysis (PCA) on selected features of each scale. Each rotated scale was inspected to ensure that the lower lamina was satisfactorily aligned to the *x-y* plane. Then we used a peak finding method to find the dense lower lamina along 1D *z*-axis phase profiles (i.e., constant *x,y*) of the 3D phase volume (i.e., electron density) obtained from PXCT (Fig 6C). Once the lower lamina was located, it was digitally translated in *z* to remove local undulations (tilt variations) (Fig 6D). This translation reduced the variations in distances between the lower lamina to upper lamina features (e.g., ridges and/or cross-ribs). This variation reduction assisted the segmentation and classification of features on the upper lamina. However, the upper lamina structures are neither continuous nor feasible to align. Hence, we needed a clustering method to classify the substructures in the upper lamina. The 1D curves were classified based on the (non-exclusive) presence of the following features: lower lamina, ridges, cross-ribs, and trabeculae. PCA dimensionality reduction (5 principal components) was carried out on 1D *z*-axis phase profiles, followed by feature vectorization using K-means clustering. We increased the number of clusters in the analysis based on the complexity and the reconstruction quality of the volume. For noisy and complex volumes, we manually labelled the K-means clusters further into ridges, cross-ribs, trabeculae and discarded obvious outliers (Fig 6E). The visual validation for our clustering is shown in Figs 3A'' - 3D'' and a cross-sectional view in Fig 6E. However, for completeness, we carried out a t-distributed Stochastic Neighbor Embedding (t-SNE) dimensionality reduction on the PCA components to visualise the inter-cluster separation (Fig S12A), and an adjusted pairwise mutual information score was calculated to evaluate the K-means clustering (Fig S12B). Although t-SNE analysis shows that the clusters were not well separated, the adjusted mutual information values, which are closer to 1, showed that our clustering was consistent and repeatable. The average 1D profiles of these upper and lower lamina features in a particular cluster are shown in Figs 3A-3D.

After classification, each of these upper and lower lamina substructures (represented by index $c$) were separately fitted to 1D Gaussians to infer the average densities ($\rho_{x,y,c}$), $z$-axis thickness ($t_{x,y,c}$), and $z$-position of these features. The Gaussian model was chosen as it accurately approximates the densities affected by the tomographic artifacts (Figs 3A, 6F). It is noteworthy that the 1D profiles dip slightly into negative values due to reconstruction artifacts from tomographic stitching. While such artifacts are common [64], in this analysis, we did not allow the fitted 1D Gaussians to be negative. Hence, these 1D Gaussian fits also reduced artifacts from tomographic stitching and noise. In the case of *B. anynana yellow* mutant scale, a double Gaussian was used for the upper lamina, instead of a single Gaussian, to account for its complex structure (tilted trabeculae beneath the ridge), in addition to a separate Gaussian fitted for the lower lamina. From these fits, the area under the Gaussians provides the phase sum along the $z$-axis ($\varphi_{x,y,c}$), and twice the standard deviation ($2\sigma_{x,y,c}$), of the Gaussians provides the thickness along the $z$-axis ($t_{x,y,c}$). We calibrated these PXCT thicknesses based on the *J. orithya* male scale's lower lamina's experimental thickness measurement (SEM and TEM) and applied the same to determine the thickness values for all the sub-structures of every scale. The phase sum and the thickness values are then used to calculate the integrated electron density along the z-axis for each pixel at the x-y plane [65]. With the electron density of a chitin monomer ($e^-_{chitin}$) and the x-ray wavelength ($\lambda_X$), we converted the phase sum into an estimate of the total number of chitin monomers. The mass density ($\rho_{x,y,c}$) of each component at a given substructure (1D profiles) was calculated by using the molecular mass of a chitin monomer ($m_{chitin}$) and Avogadro's number ($N_A$) (Eq. 1).

$$\rho_{x,y,c} = \left(\frac{\varphi_{x,y,c}}{t_{x,y,c}}\right)\left(\frac{m_{chitin}}{r_e\, \lambda_X\, N_A\, e^-_{chitin}}\right) \quad (1)$$

We emphasize that our absolute mass density estimates obtained with PXCT are based on the average ratio of electron-to-nucleon in chitin ($C_8H_{13}O_5N$) only, because the cuticular protein and pigment composition within the scales are unknown. Hence, estimates of absolute densities would increase by 11%, had we recalibrated for melanin pigment monomers instead ($C_{18}H_{10}O_2N_2$). Nevertheless, the relative differences in the average mass densities between upper and lower laminas in all four scales in Fig 4 exceed this 11% difference due to calibration choice. Consequently, the density differences between the most prominent features in the lower and upper laminas cannot be explained by a (highly unlikely) replacement of all chitin monomers entirely with melanin monomers. Since the ratio of electron-to-nucleon of a chitin monomer and a chitin polymer is identical, we used the monomer's chemical composition in our calculations for convenience.

Step-by-step procedural description of mass density estimation

1. The PXCT reconstructed volume is rotated such that the lower lamina aligns with the *x-y* plane (Fig 6B).
2. Further, a peak finding on 1D curves along the $z$-axis at every $x$-$y$ position is carried out to determine the position of the lower lamina (Fig 6C).
3. The lower lamina is digitally flattened to remove all the local undulations by aligning the peaks determined from the previous step (Fig 6D).
4. Since the upper lamina is not continuous, PCA and K-means are applied to cluster the 1D curves to segment the volume and identify the substructures of the upper lamina. We employed five PCA components and four clusters for K-means. We increased the number of clusters for scales with complex structures and validated the clustering with t-SNE (Fig 6E).
5. After segmentation, Gaussian fitting is employed on every single 1D curve of each cluster ($c$) to determine the average lower bound estimates of the mass density (Fig 6F). The Gaussian fitting ensures the estimates are more robust against the tomographic artifacts and better than the direct voxel-wise estimation (Fig S8).
6. The area under the Gaussian curve provides the phase sum along the $z$-axis ($\varphi_{x,y,c}$) (Fig 6F).
7. From TEM and SEM thickness measurements, the thickness along the $z$-axis is calibrated as twice the value of the fitted standard deviation ( $t_{x,y,c} = 2\sigma_{x,y,c}$).
8. The phase sum and the thickness can already provide the average electron density of each 1D profile.
9. Based on the assumption that the material is chitin (to convert electron to nucleon density), the mass density is estimated from the phase sum and thickness using Eq. 1.
10. The partial volume effect and the presence of melanin, which has an 11% higher nucleon-to-electron ratio than chitin, render our mass density estimates to be the lower bound.

Partial volume effect

The mass density estimations of the sub-structures should be inferred cautiously as they suffer from the partial volume effect, i.e., a limited resolution of an imaging system underestimates the density in smaller features. For example, the ectopic cuticular film of *B. anynana yellow* mutant scale is thinner than a PXCT voxel; thus, the electron mass reported in that voxel is averaged over the larger voxel size instead of the actual thickness of the upper lamina, which causes a lower density than the actual value. Though significant in the case of the thin ectopic film, all the other features (cross-ribs, ridges, trabeculae, and lower lamina) that are larger than two voxels are less affected by the partial volume effect [66]. Conservatively, all the mass density values estimated should be considered as the lower bound of the actual densities. However, the relative change in densities of sub-structures within a scale (except for the ectopic film) and densities of the same structure between different scales can be inferred with confidence since the voxel size is the same in all reconstructions. It is noteworthy that the partial volume effect can be substantially reduced by collecting PXCT data with a higher resolution (i.e., by increasing exposure times at the expense of longer data acquisition times); hence it is not an innate limitation of PXCT or the mass density estimation.

## Optical modeling

The reflectance spectrum from a scale is a collective product of a series of optical responses from the structure. Numerical methods can simulate the complete response spectrum from a given 3D structure. However, they are computationally expensive for carrying out multiple simulations that are required for parameter tuning to fit the experimental response curves. Here we used an analytical model where an incident spectrum reflected back from the lower lamina at various layers interfere to produce a response reflection spectrum. We used the scattering matrix (S-matrix) method to calculate a reflectance spectrum from the lower lamina as we can model multi-layered lower lamina ($N$ layers) with this method [67]. Scattering matrix implementation in the PyPI python library has been used, and we calculated the reflection and transmission intensity (Eq. 2) for every wavelength data point from the experiment.

$$[I_{ref}(\lambda)\ I_{trn}(\lambda)\ ] = S^{ll}\ (0, N)\ [I_{inc}(\lambda)\ 0\ ] \quad (2)$$

We implemented a *Python* fitting routine that uses our optical model to fit the wavelength-dependent refractive indices of each layer by minimizing the difference between the model output and the experimentally measured reflectance data.

We modeled a four-layer lower lamina with the layer thickness obtained from the TEM cross-section (Fig 5A). Further, we accounted for the absorbance, the numerical aperture, and the local thickness variation (obtained from the SEM measurements) to fit the Cauchy coefficients for each layer, which represent the wavelength-dependent refractive indices. The scales with reflectance and absorbance measurements were used for the modeling (Figs S9B, S9C). First, we corrected the experimental reflectance spectra to compensate for the absorbance with the corresponding absorbance measurements (Fig S9D). We defined a probability distribution of thicknesses obtained from the mean and variance of thickness from the SEM measurements to simulate the local thickness variation (Fig S9E). Then, we estimated the half-angle (30 degrees) for the cone of the incident light from the numerical aperture (0.5). Our model integrates the thin-film reflectance response from the thickness and the incident angle distribution (0 to 30 degrees) to obtain the final reflectance spectra (Fig S9F). Since the upper-lamina-removed scales were fragile, it was difficult to move them to a black tape for the reflectance measurement and bring them back to a glass slide for the absorbance measurement without damaging the scale. Hence, we measured the reflectance of the scale on top of a glass slide. To be thorough, we have also included the glass substrate background and a thin air gap (< 5 nm) between the scale and the glass slide in our model (Fig 5D). Since we used the mirror reference for the measurements and included the glass substrate in our model, we did not require to normalise the spectra nor a correction factor to account for the overestimation.

## Absorption spectrum and pigment concentration estimation

Spectrophotometer measures absorbance $A(\lambda, t)$ from the fraction of the light transmitted $T(\lambda, t)$ through a sample of thickness $t$ as a function of the wavelength of incident light $\lambda$. This absorbance is shown in Eq. (3)

$$A(\lambda, t) = -\log_{10} T(\lambda, t) \ . \quad (3)$$

In the case of butterfly scales, this absorbance is affected by the *types* and *density* of light-absorbing pigments within the scale, as well as the scale's thickness. Hence, to compare the density of pigments within different regions of a scale (i.e., upper versus lower lamina) and between similar scales, we need to account for the corresponding thicknesses of pigment-containing features within the scale. Fortunately for us, this thickness information can be inferred from SEM and PXCT measurements.

The SEM cross sections of the four types of scales presented in this manuscript each show relatively little variations in the thickness of their lower lamina (Figs S5, S6). Unlike the lower lamina, however, the upper lamina of each of these four scale types contain a complex structure with ridges, cross-ribs, trabeculae and window regions. Hence, the probability distribution of thicknesses across such a complex structure should be estimated for its transmittance model. For each scale, we estimated the thickness distribution of its upper lamina from 100 randomly sampled 2 μm-diameter patches from its corresponding PXCT reconstruction volume. This diameter was chosen to match the illumination spot sizes using in the micro-spectrophotometry measurements (described in an earlier section in the methods).

Overall, the average transmittance for such an upper and lower lamina structure can be modeled using the following pair of Beer-Lambert law:

$$T_{U+L}(\lambda) = \mathrm{e}^{-(\mu_L(\lambda) t_L)} \sum_{t_U} \Pr(t_U) \mathrm{e}^{-(\mu_U(\lambda) t_U)} \quad (4)$$

$$T_L(\lambda) = \mathrm{e}^{-(\mu_L(\lambda) t_L)} \quad (5)$$

where $\mu_L(\lambda)$, $\mu_U(\lambda)$ and $t_L$, $t_U$ are pairs of absorption spectra and thickness of the lower and upper lamina respectively. $\Pr(t_U)$ is the probability distribution of the upper lamina thickness (from PXCT reconstruction), which allows us to account for the varying thicknesses of the upper lamina features in each scale. These thickness values $t_L$ were calibrated based on the thickness of the lower lamina experimentally measured from the SEM, which is more precise than that from PXCT.

To reduce the partial volume effect from upper lamina feature thicknesses ($t_U$) inferred from PXCT, we found the scaling factor needed to match the lower lamina thicknesses between PXCT and SEM measurements for each scale. This butterfly-scale-dependent scaling factor was used to rescale each scale's upper lamina thickness distribution to minimize the partial volume effect.

Here's what we did, in detail. For each scale, whose lower-lamina thickness $t_L$ was measured via SEM, we fitted its measured lower-lamina-only absorbance $T_L(\lambda)$ to Eq. (5) to find the absorption spectra of the lower lamina $\mu_L(\lambda)$. These fitted parameters were substituted into Eq. (4), together with the upper lamina thickness distribution $Pr(t_U)$ from PXCT (after correcting for partial volume effects) and fitted to the upper-plus-lower lamina microphotospectra absorbance $T_{U+L}(\lambda)$. Doing this gave us the absorption spectra of just the upper lamina $\mu_U(\lambda)$.

Notice that now that we have the thickness-independent absorption spectra of the upper and lower laminas (i.e., $\mu_U(\lambda)$ have $\mu_L(\lambda)$), we can determine the chemical signature and density of the pigments in each. To facilitate this, we decomposed the absorption spectrum $\mu(\lambda)$ into two independent parts:

$$\mu(\lambda) = c\varepsilon(\lambda) \tag{6}$$

where $c$ is the molar concentration of the light-absorbing pigments and $\varepsilon(\lambda)$ is the molar absorption spectrum of these pigments. Should the upper and lower laminas have comparable wavelength dependence of $\varepsilon(\lambda)$ (i.e., $\varepsilon_U(\lambda)/\varepsilon_L(\lambda)$ is a wavelength-independent constant) we can be more confident in asserting that they contain the same type of pigments, except at different relative concentrations. Fig S2 shows the molar absorption spectrum $\varepsilon(\lambda)$ for the upper and lower lamina of four scales studied here (that of the lower lamina of *J. orithya* male was omitted because it was too noisy). Fig S2 also compares the relative molar concentrations of pigments in these scales.

Transmission electron microscopy of resin-embedded scales

Small pieces of each wing section were cut and fixed in 2.5% Glutaraldehyde in (Phosphate-buffered saline) PBS overnight at 4°C followed by three washes in PBS. Samples were then processed at the Electron Microscopy Unit at the Yong Loo Lin School of Medicine, NUS. Briefly, samples were post-fixed in 1 or 2% OsO4, pH7.4 for 1 hour at room temperature under a fume hood and washed in deionized water for 10 minutes for 2 changes at room temperature. Dehydration was through an ascending ethanol series at room temperature: 25% ethanol for 5 minutes, 50% and 75% ethanol for 10 minutes, 95% and 100% ethanol for 20 minutes. Samples were then transferred to 100% acetone for 20 minutes, 2 times. Infiltration was with the following steps: a) 100% acetone : Araldite resin (3:1) for 30 mins at room temperature b) 100% acetone : Araldite resin (1:1) for 1 hour at room temperature c) 100% acetone : Araldite resin (1:6) for overnight at room temperature d) Pure Araldite resin for 30 minutes at room temperature then for 30 minutes in a 40°C oven e) Change to pure Araldite resin again for 1 hour in a 45°C oven f) Last change to pure Araldite resin for 1 hour in a 50°C oven. Finally, samples were embedded in fresh resin and allowed to polymerize for 24 hours in a 60°C oven. For some *J. orithya* male samples, no post-fixation in OsO4 was carried out.

Ultrathin sections were cut using a Leica Ultracut UCT Ultramicrotome (Leica Microsystems, Germany) and transferred to formvar coated grids. For the *J. orithya* samples grids were either left unstained or stained with lead citrate. For the *B. anynana* samples, all grids were stained with lead citrate. Images were acquired on a FEI Technai T12 Transmission Electron Microscopy (FEI, Thermo Fisher Scientific, U.S.A).

## Data availability

Source data for all reflectance, absorbance, and thickness measurements are provided with this paper in Supplementary Data 1 and Supplementary Data 2 (Excel files). Cropped PXCT volumes used for segmentation and density analysis are openly available in Zenodo (10.5281/zenodo.12703866) [68]. The full reconstructed PXCT volumes and corresponding raw data are not publicly hosted due to file-size constraints; they are available upon request via email to the corresponding author.

## Code availability

The custom codes (Jupyter notebooks) for X-ray ptycho-tomography 3D volumetric data analysis and optical modeling are available in the same Zenodo repository (10.5281/zenodo.12703866) [68].

## Acknowledgments

We thank Ms. Tan Suat Hoon and Mr. Sim Aik Yong of the Electron Microscopy Unit at the Yong Loo Lin School of Medicine for their service in preparing the TEM resin blocks and sections. We acknowledge Diamond Light Source for time on I13-1 under proposal MG23967-1. We thank Dr. Darren Batey and Dr. Silvia Cipiccia for technical support and Dr. Simone Sala for help with data processing during the PXCT measurements taken at the I13-1 beamline. We would also like to acknowledge the use of the High Performance Computing facilities at the Diamond Light Source for the processing of the PXCT data. This work was supported by National Research Foundation (NRF) Singapore, Competitive Research Program grants NRF-CRP20-2017-0001(AM), NRF-CRP25-2020-0001 (AM), Investigatorship award NRF-NRFI05-2019-0006 (AM), National University of Singapore Early Career Research Award (NDL), and European Research Council Consolidator Grant S-BaXIT (866026) (PT).

## Author contributions

Conceptualization: AP, BJD, AM, NDL
Methodology: DB, AP, BJD, PT, AM, NDL
Formal analysis: DB, BJD, YCL, ZS, NDL
Investigation: DB, AP, BJD, CF, NDL
Visualization: DB, AP, CF
Funding acquisition: AM, NDL, PT
Project administration: AM, NDL
Supervision: AM, NDL
Writing – original draft: DB, AP, BJD
Writing – review & editing: DB, AP, BJD, CF, PT, AM, NDL

## Competing interests

Authors declare that they have no competing interests.

## Figures

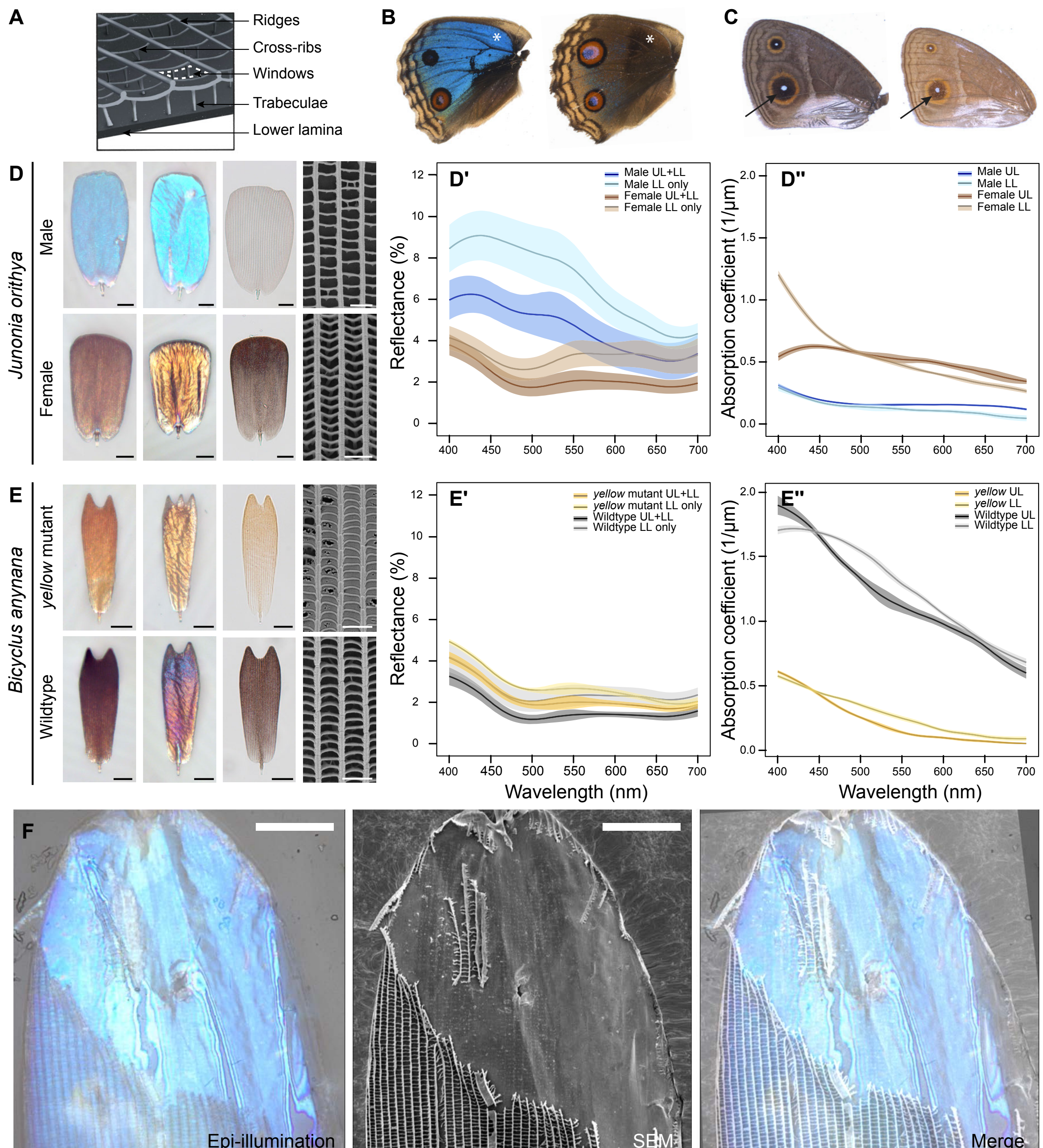

**Fig 1. Reflectance and absorption profiles of *Junonia orithya* and *Bicyclus anynana* scales.** (A) Schematic of a wing scale showing a typical scale structure (digitally illustrated in Procreate). (B) Dorsal hindwings of a *J. orithya* male (left) and female (right). White stars indicate the regions sampled. (C) Ventral forewings of a *B. anynana* wildtype (left) and *yellow* mutant (right). Black arrows indicate the regions sampled. (D, E) Optical and SEM images of a male blue scale and a female brown scale of *J. orithya* (D), and a *yellow* mutant scale and wildtype scale of *B. anynana* (E). First column: Abwing images; Second column: Adwing images; Third column: Optical microscopy images of scales immersed in a chitin refractive index matching liquid that eliminates structural colors and measures pigmentary colors; The interference pattern in *J. orithya* male scale immersed in a chitin refractive index matching liquid is caused by an imaging artifact called color Moire. This is caused due to downsampling an image with a periodic structure to a smaller number of pixels in Fig 1. Fourth column: SEM

images of the upper lamina structures. Scale bars for the optical images are 20 μm and the SEM images are 2 μm. (D′, E′) Reflectance spectra (Solid curve = mean spectrum; shaded band = ± 1 SD (standard deviation)) of abwing *J. orithya* (D′) and *B. anynana* (E′) scales with intact upper and lower laminas (UL+LL) and scales with the upper laminas removed physically (LL only). (D″, E″) Estimated absorption coefficient spectra of *J. orithya* (D″) and *B. anynana* (E″) scales for their upper laminas (UL only) and lower laminas (LL only). At least three scales have been imaged from 2 different individuals; the exact sample sizes and replicate types for each curve are detailed in the 'UV-VIS-NIR microspectrophotometry' subsection of Methods. Source data for panels D′, E′, D″, and E″ are provided in Supplementary Data 1. (F) An illustrative example of a *J. orithya* blue scale with the upper lamina removed, imaged under epi-illumination, scanning electron microscopy (SEM), and a merge of the two. The scale bars in (F) are 20 μm.

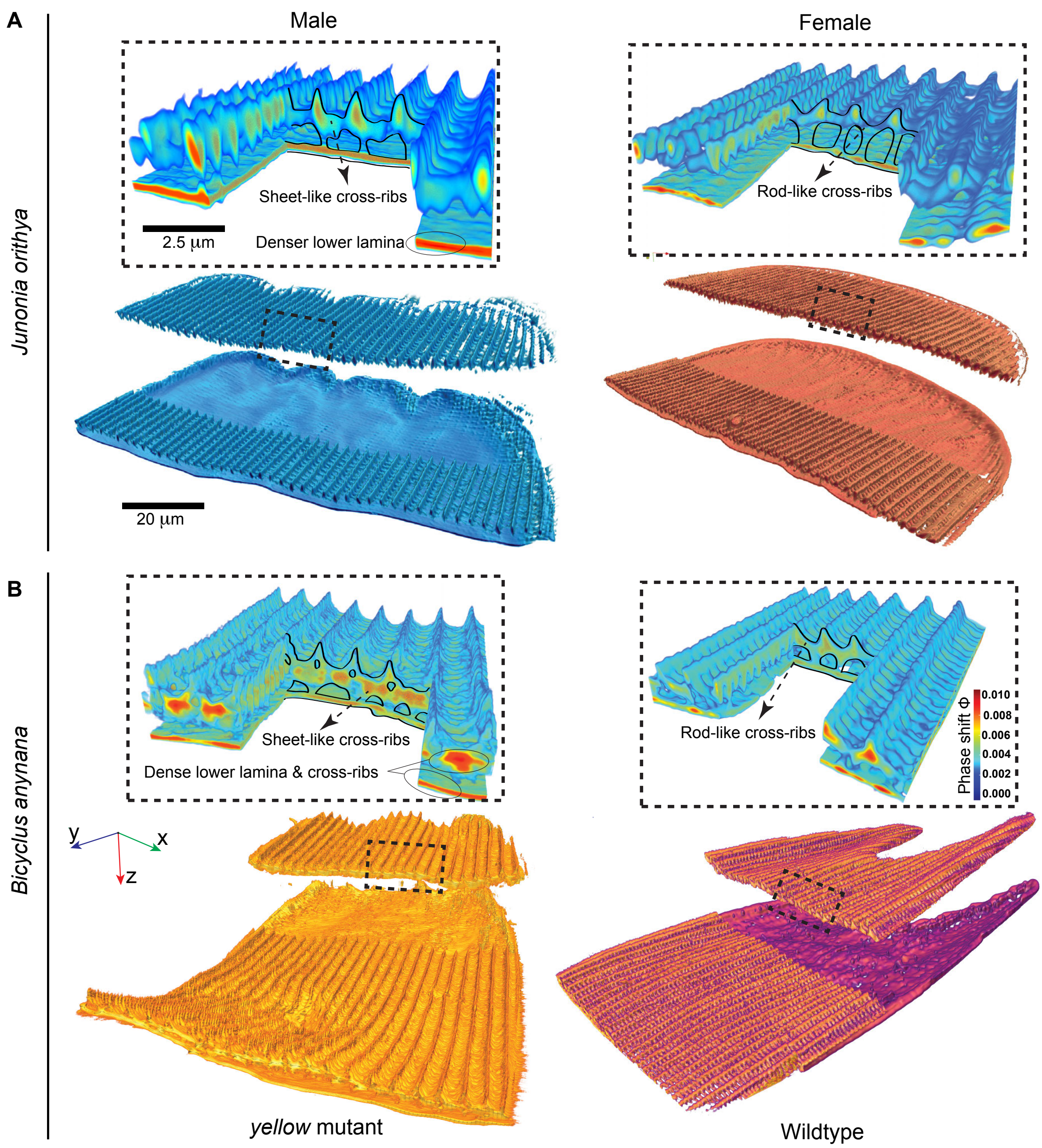

**Fig 2. Volume segmentation of 3D PXCT reconstructions of *J. orithya* and *B. anynana* scales.** False color density contour renders of a (A) *J. orithya* male blue scale (left), and a female brown scale (right); and (B) a *B. anynana yellow* mutant scale (left) and a wildtype scale (right). The renders show digitally separated upper laminas cut out from the rest of the scales. Black, dashed rectangles indicate the regions magnified for 3D electron density maps of the upper and lower lamina. Cross-sections along the ridges and across the ridges in the 3D heat maps show electron density variations within the scales. Annotated black lines outline the structural differences across scales. The figure shows that the PXCT reconstructions provide 1) 3D structural data over a large field of view, 2) nano-scale density information of the specimen, and 3) digital segmentation of the reconstructed volume to estimate density variations of each sub-structure in a physically intact scale.

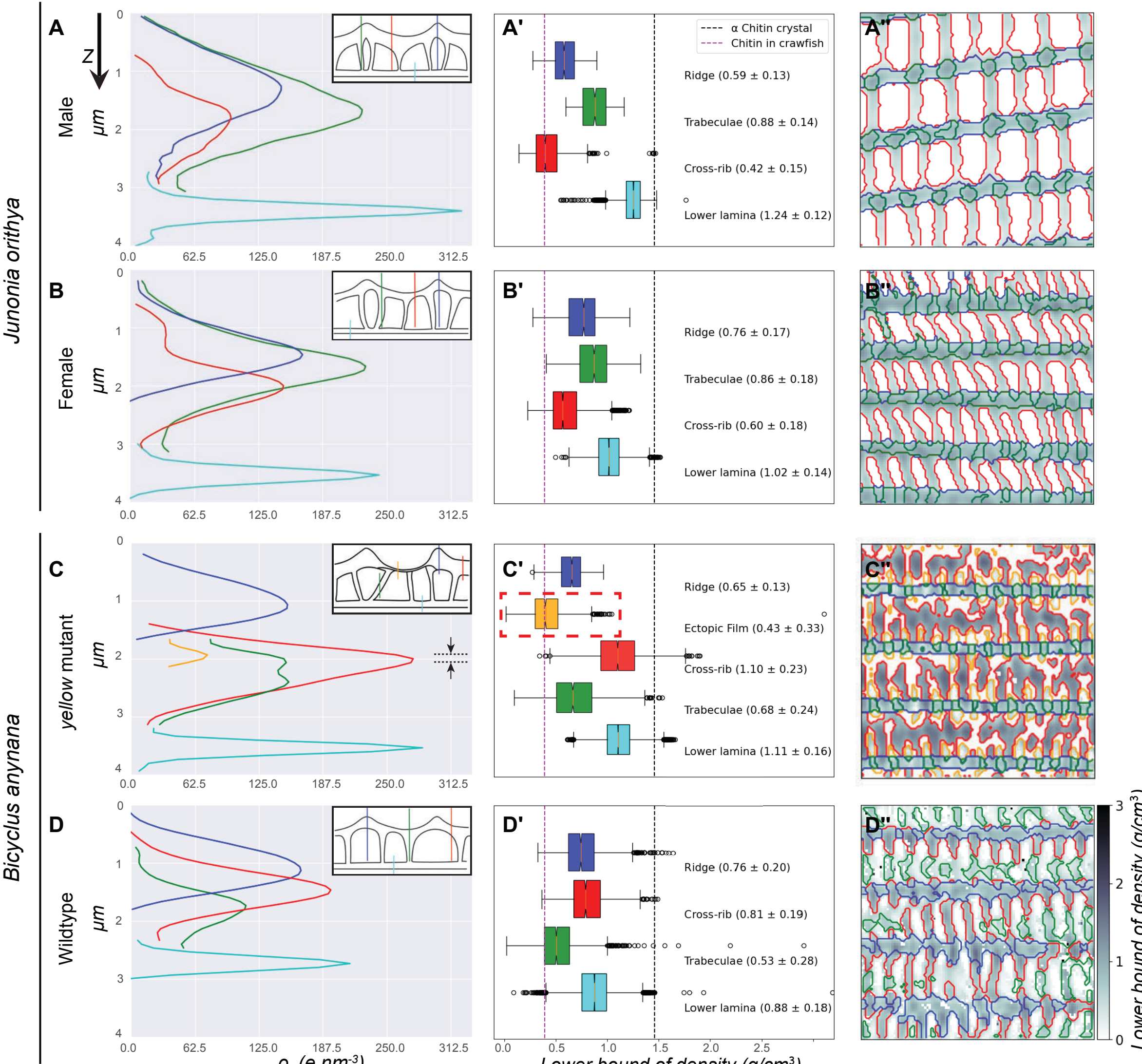

**Fig 3. Mass density estimations from the segmented volumes of *Junonia orithya* and *Bicyclus anynana* scales.** Sub-volumes from the PXCT reconstructions are segmented and clustered based on mass distribution (details in the step-by-step procedural description of the mass density estimation in Methods). (A-D) The mean electron density distribution along the z-axis (from abwing (upper) to adwing (lower)) of different upper lamina sub-structures with respect to the lower lamina of *J. orithya* male and female scales (A, B) and *B. anynana yellow* mutant and wildtype scales (C, D). Insets show a schematic view of the scales with colored lines marking the transected regions measured by the corresponding-colored curves. Cyan – lower lamina, red – cross-ribs, green – trabeculae, blue – ridges, and yellow – ectopic film . We can see from the insets that the lower lamina is further away (in the z-axis) from other upper lamina structures, which is why the cyan curves are well separated and do not overlap with the other curves. (A′-D′) Box plots (Centre line = median; box = 25th–75th percentiles; whiskers = values within 1.5 × IQR; points = outliers) of the estimated mass densities clustered as lower lamina, cross-ribs, ridges, and trabeculae reveal density variations across the scale sub-structures within each pair of differentially pigmented scales. Mass densities of α-chitin and chitin in crawfish are plotted to show the range of chitin densities reported in various organisms. The red dotted rectangle over the ectopic film density box plot in C′ indicates that the density estimate is unreliable, as the ectopic layer thickness is significantly smaller than the voxel-resolution limit of the PXCT. Statistics are derived from one complete PXCT tomogram per scale type (one biological replicate), and 10,000 independent voxel-column measurements within the PXCT volume (technical replicates) are used to compute the density box plots. (A″-D″) 2D mass density maps (upper lamina only) of the sub-volumes identified with colored contours. Source data for this figure are available in Zenodo (10.5281/zenodo.12703866).

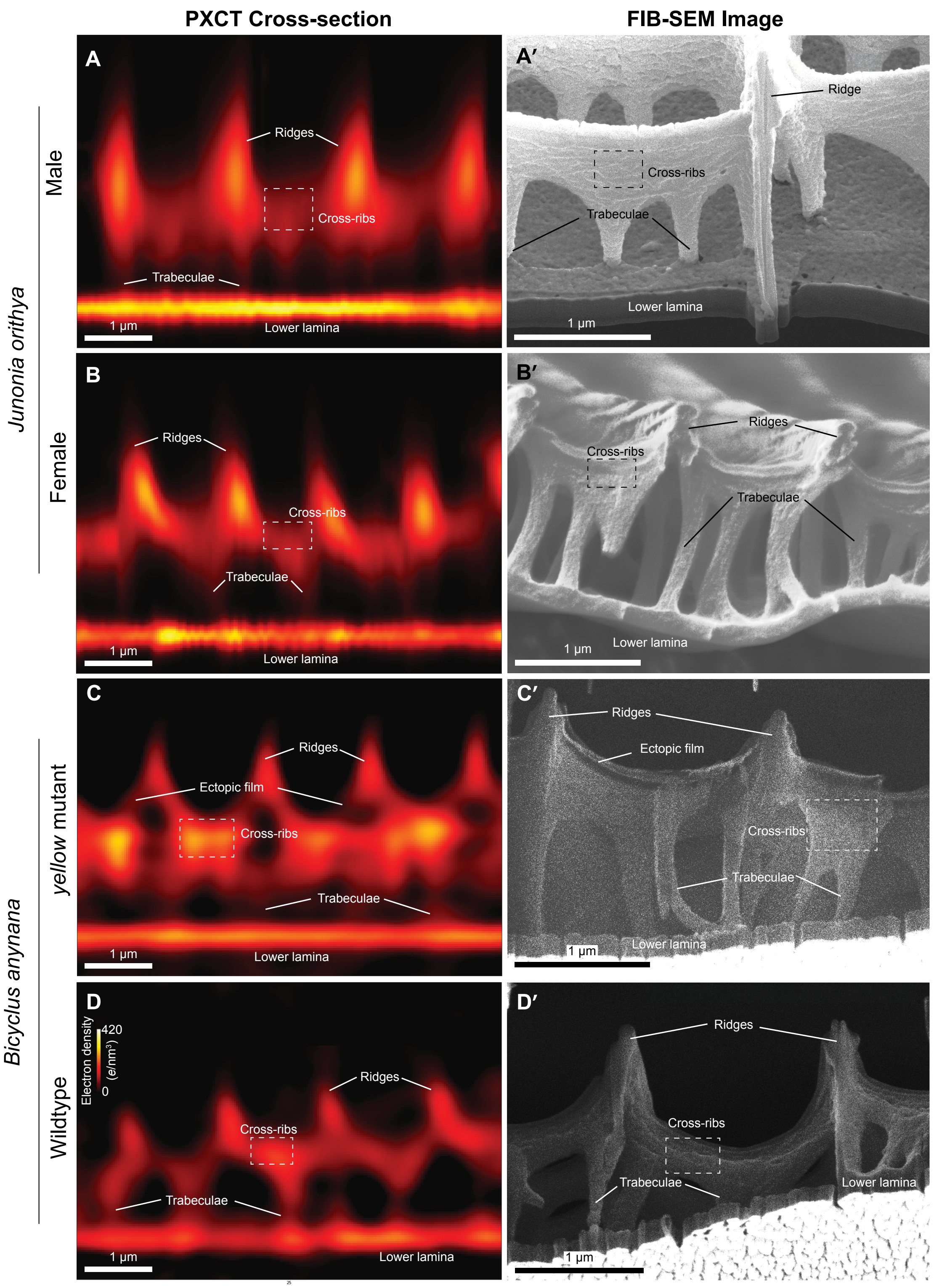

**Fig 4. Comparison of PXCT cross-sections with the FIB-SEM images of *Junonia orithya* and *Bicyclus anynana* scales.** (A-D) Cross-sections of the scales that are digitally cut from the PXCT reconstructions show the resolved structures and density distributions within the sub-structures. One PXCT tomogram per scale type. (A'-D') FIB-SEM cross-sections of similar scales of each type. The traditional FIB-SEM images show structural similarity to the PXCT reconstructions. For *J. orithya*, 3-5 scales were imaged across 3 individuals. For *B. anynana*, 4 scales were imaged across 2 individuals.

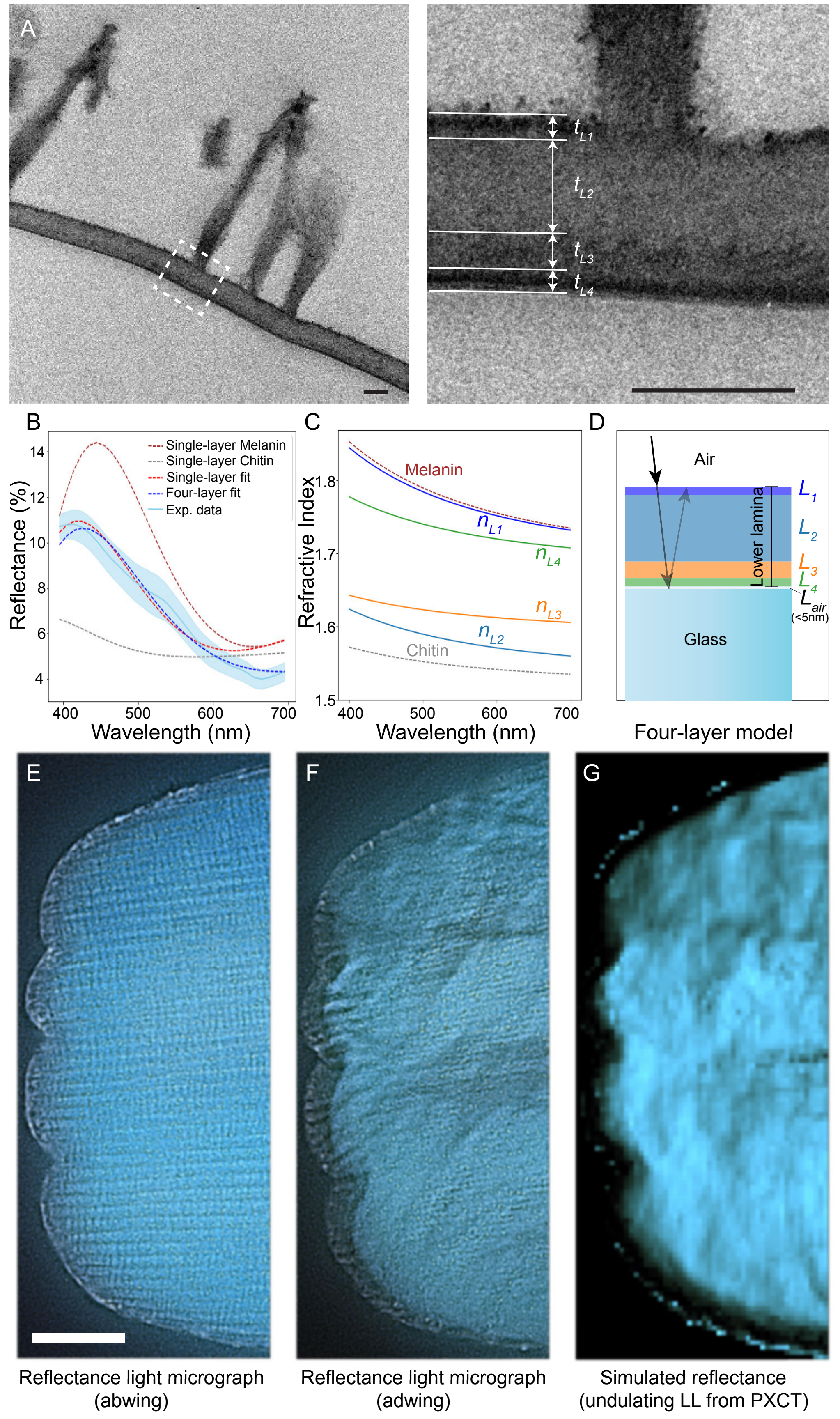

**Fig 5. A multi-layer thin-film interference model fit for reflectance spectra of *Junonia orithya* male blue scale.** (A) TEM image of a blue scale of *J. orithya* male. The sample has not been stained with uranyl acetate or lead citrate; only osmium tetroxide is used for the postfixation. The white rectangle indicates the region cropped for the magnified view. The lower lamina shows two brighter layers sandwiched between the darker layers. The scale bar is 200 nm. The same four-layer contrast was observed in 12 images collected from 3 *J. orithya* male scales from 1 individual; layer thicknesses were quantified from the image shown here. (B) Experimentally measured reflectance spectrum of a *J. orithya* male scale (the upper lamina physically removed) placed on top of a glass slide, shown in light blue; solid curve represents the mean spectrum (shaded band = ± 1 SD) of six measurements from n = 2 independent scales. This spectrum is superimposed with four simulated spectra: a single 190 nm-thick layer of chitin placed on a glass slide using chitin's refractive index from unpigmented glassy scales; a single 190 nm-thick layer of melanin on glass using melanin's refractive index determined from bird barbules; a four-layer 190 nm-thick film on glass where the refractive indices in each layer are allowed to vary, and taking layer thicknesses from (A); a single 190 nm-thick layer thin film whose refractive indices are allowed to vary. The models account for a thin air gap with variable thickness; however, fitting results suggest its contribution is negligible. Source data for this panel are provided in Supplementary Data 1. (C) The wavelength-dependent refractive indices (Cauchy coefficients) of melanin, chitin, and the corresponding best-fit indices of the four layers of the lower lamina in the four-layer model in (B). (D) Schematic of four-layer model: four-layered lower lamina ($L_1$, $L_2$, $L_3$, and $L_4$) on top of a glass slide with a thin layer of air ($L_{air}$) between the lower lamina and the glass. (E, F) Epi-illumination light microscopy images of the abwing side (E) and the adwing side (F) of the same *J. orithya* male blue scale imaged with PXCT. The scale bar in panel E represents 20 μm and applies to panels (E-G). (G) A reflectance map was simulated using the four-layer model and an estimated tilt map of the lower lamina from the PXCT reconstruction. The local tilt affects the color and intensity response that correlates well with the adwing light microscopy image (E).

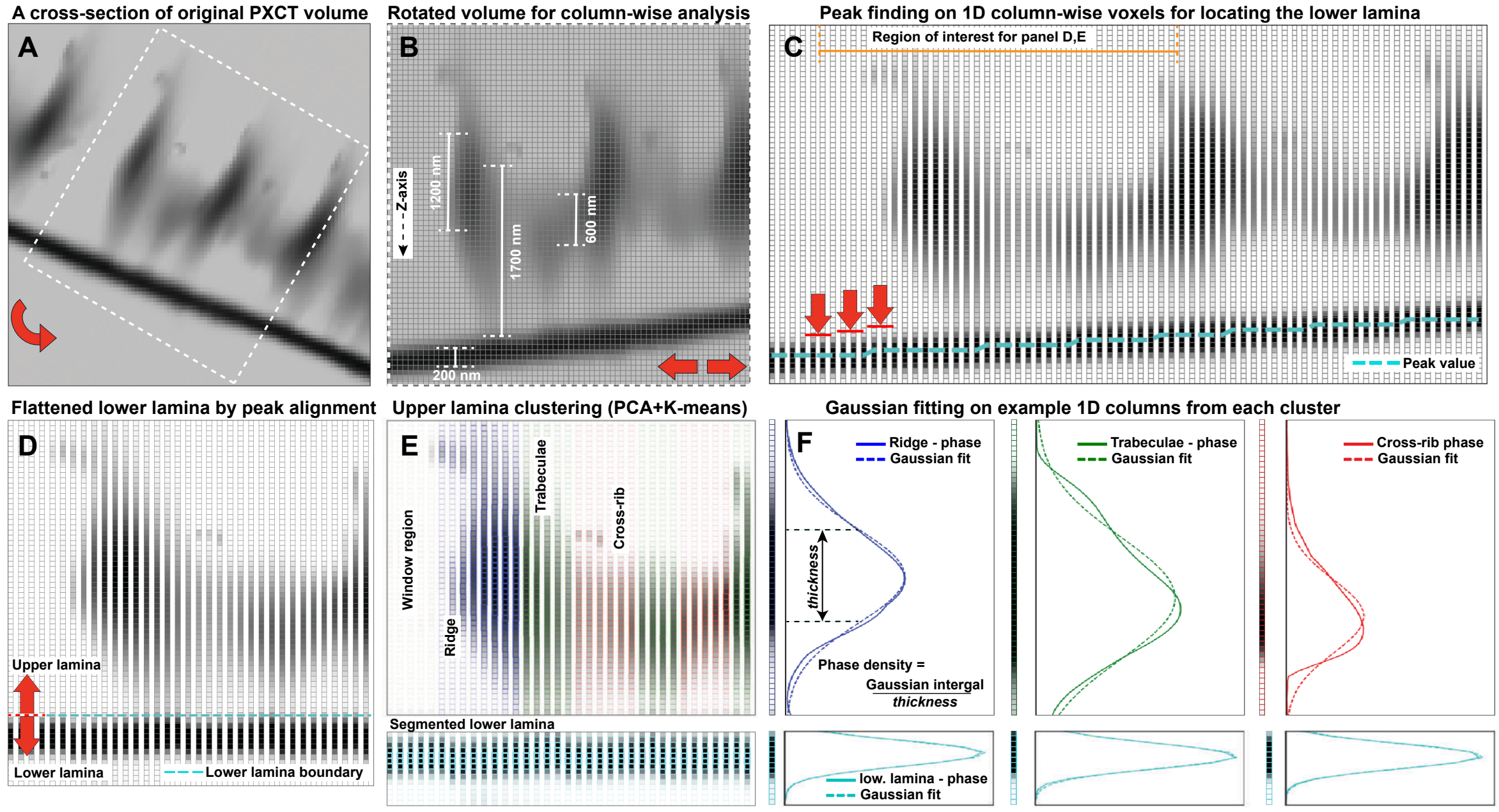

**Fig 6. *PXCT density estimation schematic.*** (A) A 2D cross-section of the PXCT reconstruction volume of a blue scale of *J. orithya* male; the boxed region is selected for the demonstration. Each voxel's reconstructed phase density is shown in grayscale (darker means denser). Red arrows in each panel indicate the next key transformation step. In this panel, the red arrow shows the rotation direction of the volume (B). The volume is rotated by the global mean tilt so that the lower lamina of the entire scale (which extends beyond this field of view) is, as much as possible, aligned with the *x*-*y* plane. Consequently, the trabeculae are conveniently aligned along the vertical *z*-axis for column-wise analysis. Red arrows indicate that the volume will be laterally partitioned into multiple 1D (*z*-axis) voxel columns next. (C) Using peak-finding, we separately detect the position of the lower lamina in each 1D column (indicated by cyan markers). Red arrows indicate that these detected peaks will be computationally aligned in the next step, since we cannot remove the local undulations in the lower lamina of the entire scale with a single rotation. (D) Local undulations in the lower lamina are flattened via peak alignment. Red arrows indicate that the lower lamina will be segmented from the upper lamina features within each 1D column in the next step. (E) The upper lamina feature vectors are reduced and clustered into four categories—window, ridge, trabeculae, and cross-rib — using PCA followed by K-means clustering. (F) The 1D profiles of the lower lamina and the upper lamina features of each column are then fitted to Gaussian functions to reduce partial volume effects and estimate thickness and phase density. This fitting is separately done on every 1D column throughout the volume. Shown here are examples from one column per cluster (ridge, trabeculae, and cross-rib).

# Supplementary Information for "Nanoscale cuticle mass density variations influenced by pigmentation in butterfly wing scales."

Deepan Balakrishnan[1,2†], Anupama Prakash[1†§], Benedikt J. Daurer[3†], Cédric Finet[1], Ying Chen Lim[2,4], Zhou Shen[2,4], Pierre Thibault[5], Antónia Monteiro[1*], and N. Duane Loh[1,2,4*]

[1]Department of Biological Sciences, National University of Singapore, Singapore.
[2]Center for Bioimaging Sciences, National University of Singapore, Singapore 117557.
[3]Diamond Light Source, Harwell Science & Innovation Campus, Didcot OX11 0DE, UK.
[4]Department of Physics, National University of Singapore, Singapore 117551.
[5]Università degli Studi di Trieste, Trieste 34127, Italy.
[†]These authors contributed equally to this work.
[§]Current address: Centre for Engineering Biology and Department of Bioengineering, Imperial College London, SW7 2AZ, UK.
*Corresponding authors. Email: duaneloh@nus.edu.sg, antonia.monteiro@nus.edu.sg

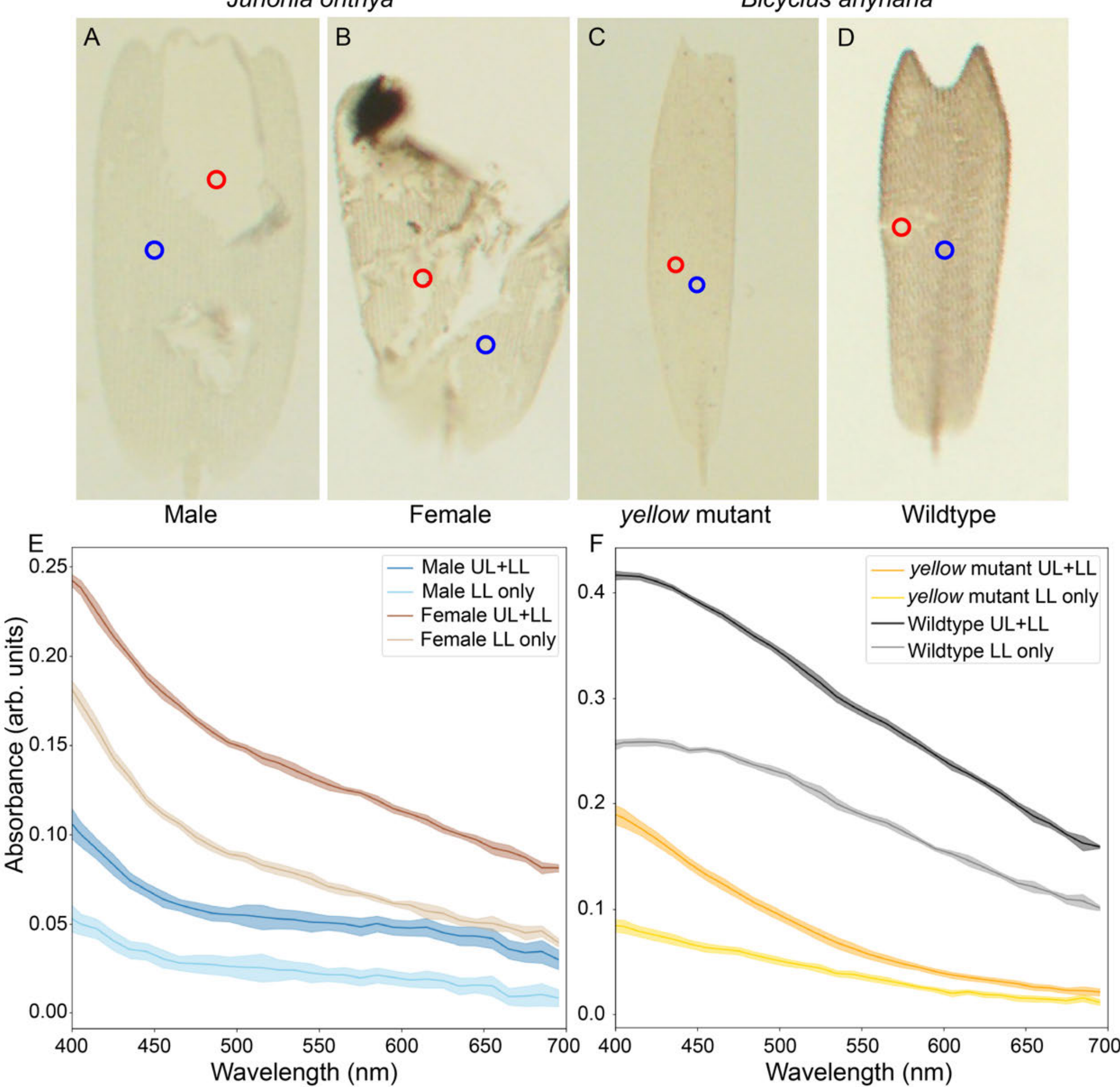

**Fig S1. Optical microscopy images and absorbance spectra of refractive index matched *Junonia orithya* and *Bicyclus anynana* scales** (A-D) Illustrative images of a *Junonia orithya* male blue scale (A) and female brown scale (B), and *Bicyclus anynana yellow* mutant scale (C) and wildtype scale (D) immersed in a refractive index matching fluid (clove oil). The images show partial removal of the upper lamina in all scale types. Upper laminas were easier to remove in the blue scales, with lower success in the *Bicyclus* scales. Immersion in a refractive index matching fluid removes the influence of structural colors and highlights the presence of pigments. The blue and red circles represent approximately the regions for spectrophotometer measurement with intact upper and lower laminas (UL+LL) and with the upper laminas removed physically (LL only) respectively. (E, F) Absorbance spectra of the four different scale types with intact upper and lower laminas (UL+LL) and scales with the upper laminas removed physically (LL only). Solid curve = mean spectrum; shaded band = ± 1 SD (standard deviation). Exact sample sizes and replicate types for each curve are detailed in the 'UV-VIS-NIR microspectrophotometry' subsection of Methods. Source data are provided in Supplementary Data 1.

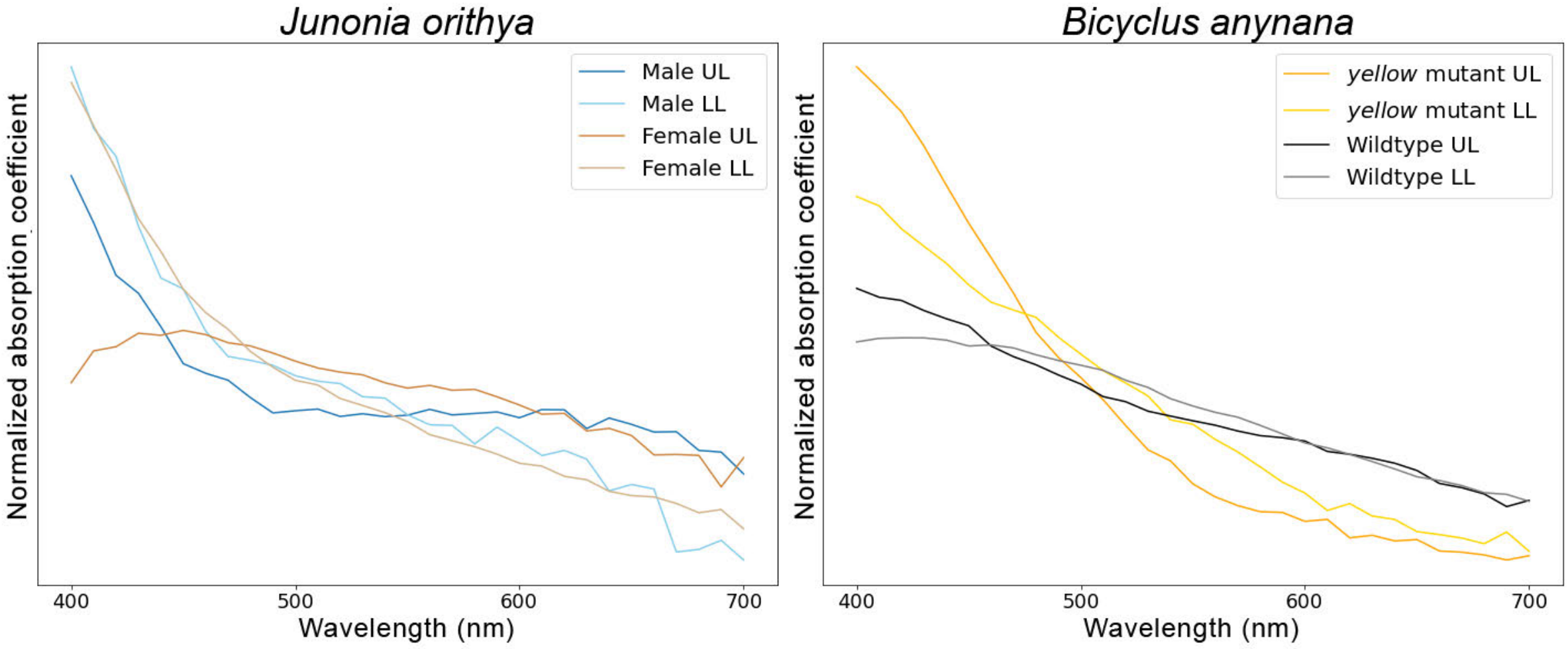

**Fig. S2. Chemical signature and relative pigment density from the absorption spectra of upper and lower laminas of *Junonia orithya* and *Bicyclus anynana*.** Absorption spectra are area normalized, i.e., the sum of the spectra were normalized to be uniform for all the spectra, to facilitate comparison of spectral shapes, which correspond to different chemical signatures of *Junonia orithya* (left) and *Bicyclus anynana* (right). Area normalization is chosen over other methods as it highlights the relative changes between the spectra across all wavelengths. The corresponding normalization factors for the curves are as follows: JM UL = 5.30, JM LL = 4.02, JF UL = 15.82, JF LL = 16.80, BM UL = 6.90, BM LL = 8.51, BW UL = 36.92, and BW LL = 39.38. When the chemical signatures of the scales are identical, the normalization factors represent relative pigment densities. Based on these profiles, we hypothesize that the *B. anynana* wildtype black scale is a complex mixture of various melanins [1]. In the *yellow* mutant scale, the loss of Dopa-melanin changes the absorbance profile and visible color of scale [1]. Less is known about the pigmentation in *J. orithya*. The absorption profile of the *J. orithya* female UL resembles the *B. anynana* wildtype UL. This might suggest that the *J. orithya* brown color is also a complex mixture of various pigments, possibly a higher concentration of tryptophan-derived ommochromes with some melanins [2]. Knockouts of candidate melanin and ommochrome pathway genes in this species will be required to test this hypothesis. JM: *Junonia* male; JF: *Junonia* female; BW: *Bicyclus* wildtype; BM: *Bicyclus yellow* mutant. UL: upper lamin; LL: lower lamina;

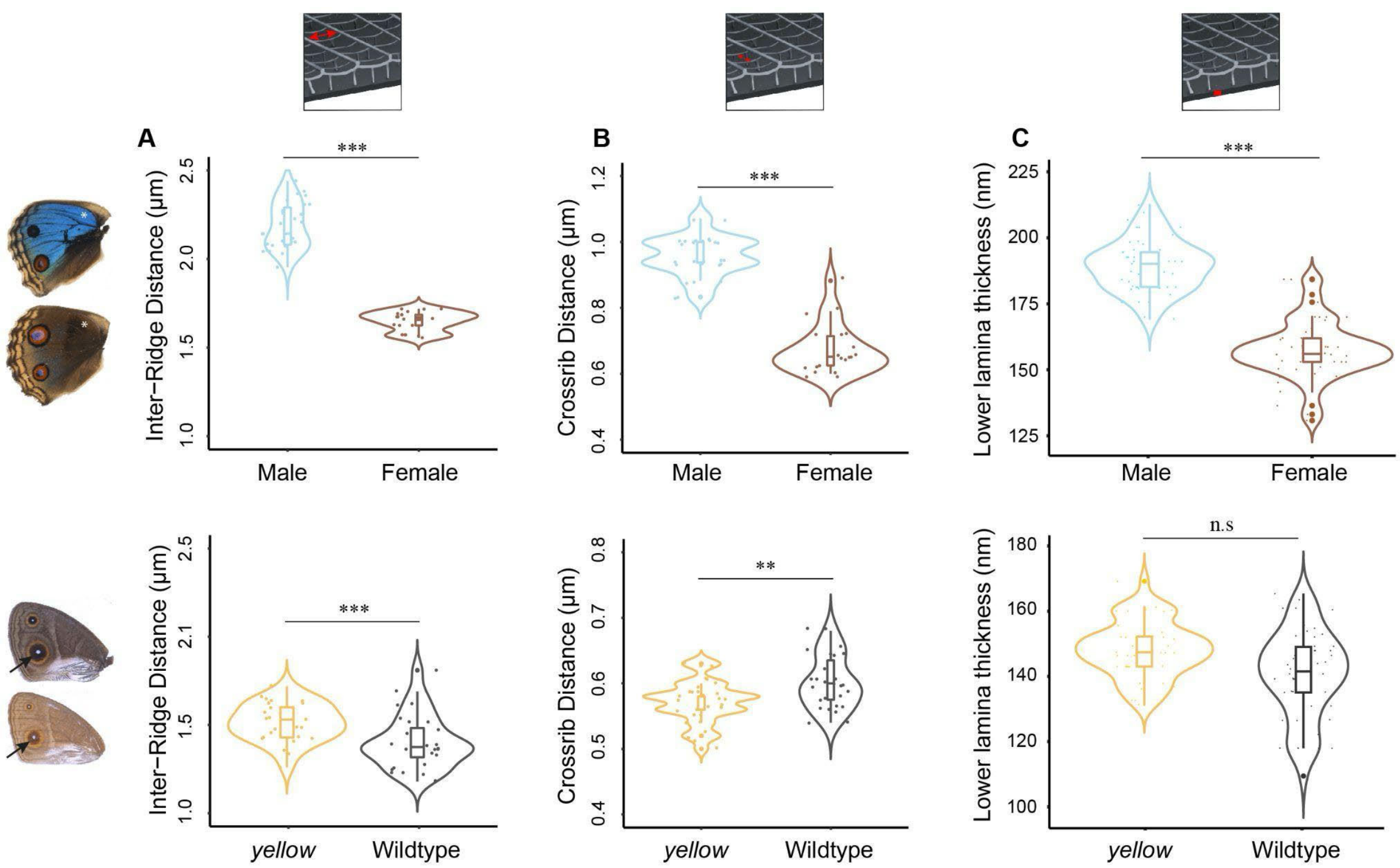

**Fig. S3. Scale structure measurements of *Junonia orithya* and *Bicyclus anynana* scales.** Violin plots of the inter-ridge distance (A), cross-rib spacing (B), and the thickness of the lower laminas (C) of male and female *J. orithya* scales (top row) and the *yellow* mutant and wildtype *B. anynana* scales (bottom row). All boxplots show the median, lower and upper quartiles, and whiskers up to 1.5 times the inter-quartile range. Asterisks indicate significant differences: $**p \leq 0.01$, $***p \leq 0.001$. Three to six scales were measured across five individuals (n=5) for each scale type. The schematics at the top depict a typical scale structure and the parameters measured (marked in red), which are digitally illustrated in Procreate.

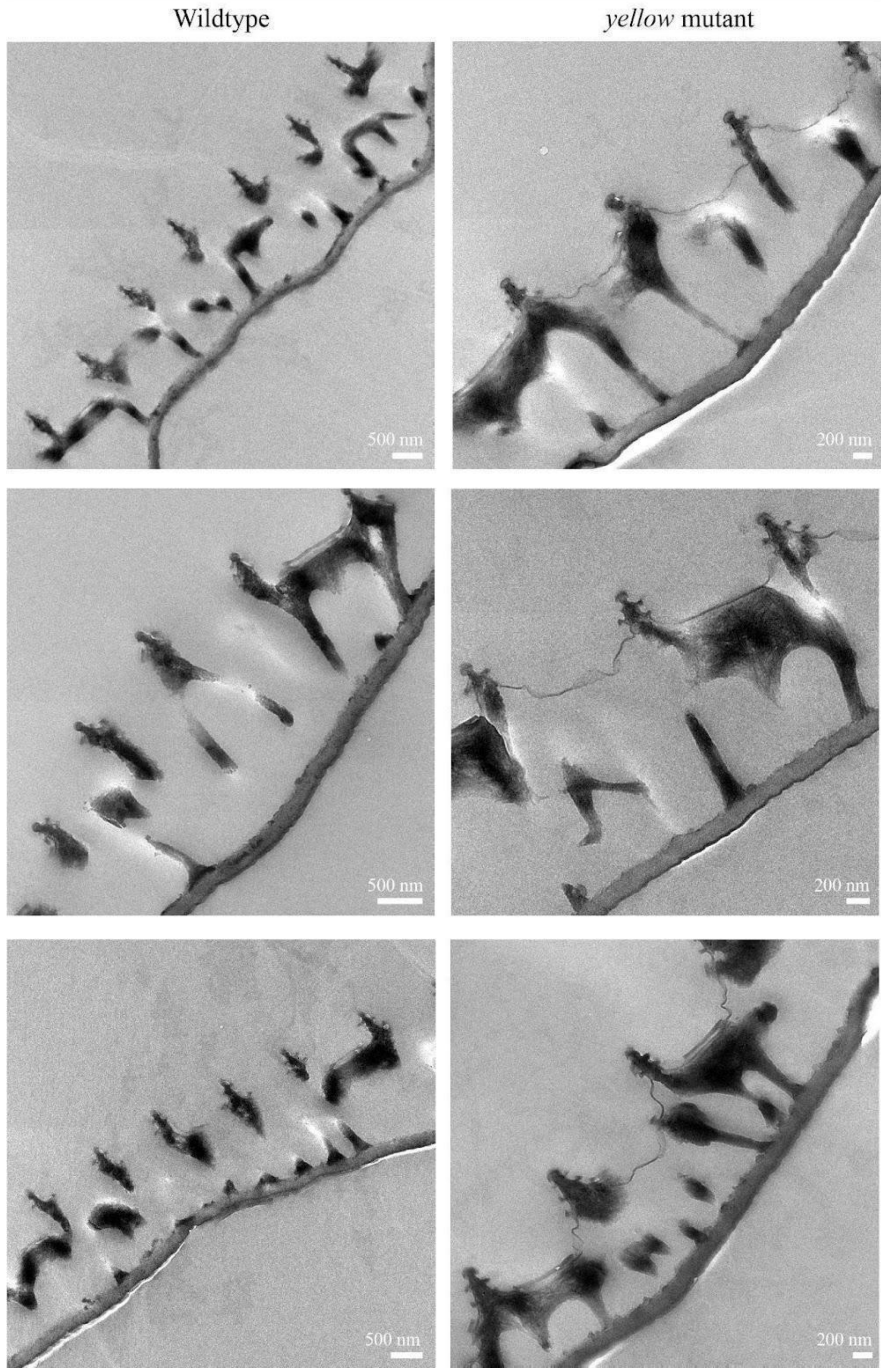

**Fig. S4. TEM images of *Bicyclus anynana* wildtype and *yellow* mutant scale cross-sections.** The *yellow* mutant scales have a 10-20 nm thick ectopic cuticular film on the upper lamina, closing the windows and more extended sheet-like cross-ribs. Representative TEM images are shown here, and TEM images of six scales were captured for each scale type.

*Junonia orithya*

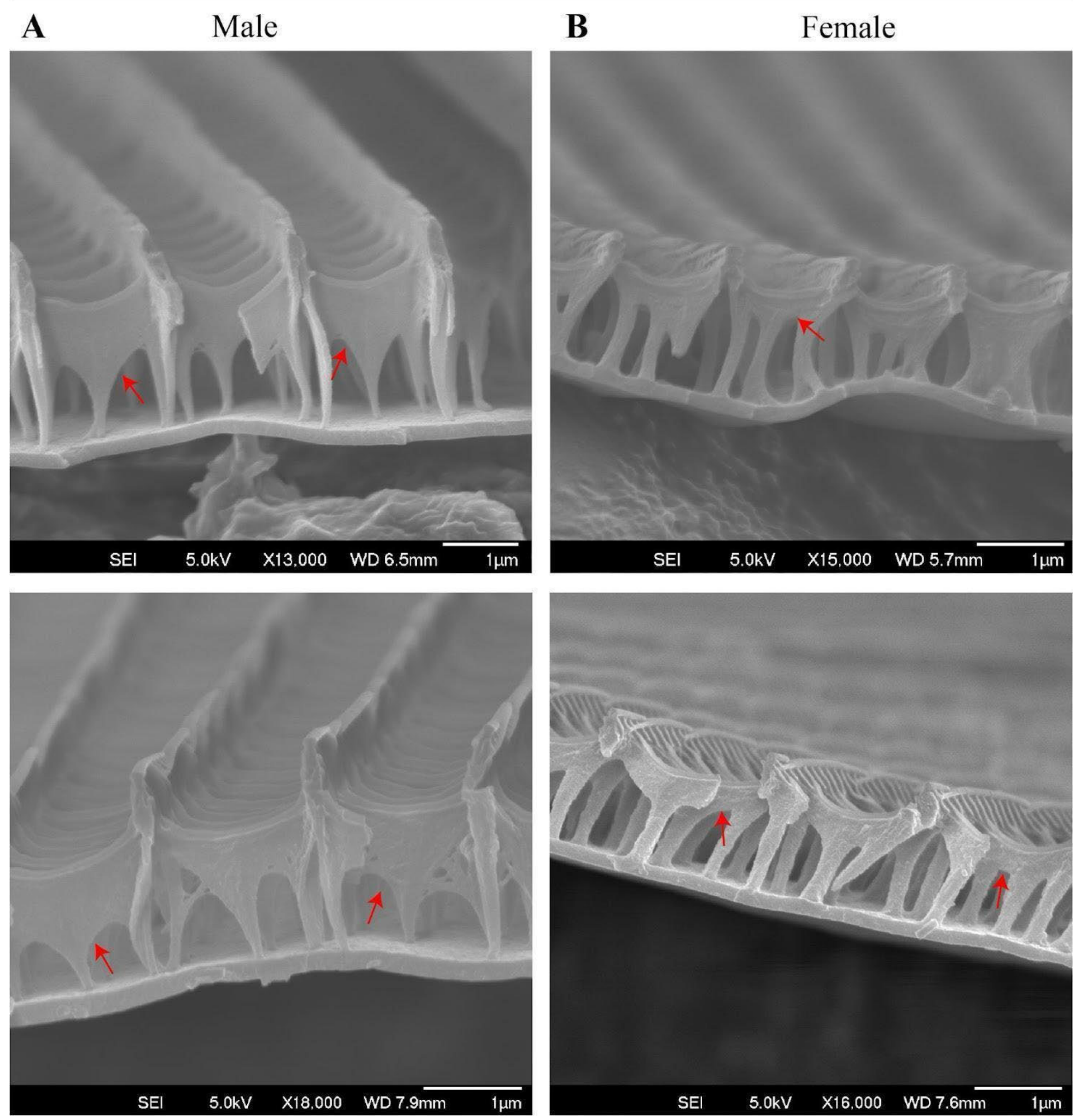

**Fig. S5. Freeze-fractured SEM cross-sections of *Junonia orithya* male and female scales.** (A), Red arrows indicate the sheet-like extensions of the cross-ribs in male scales. (B) In the female scales, the extensions of the cross-ribs (indicated by red arrows) are less prominent compared to the male scales.

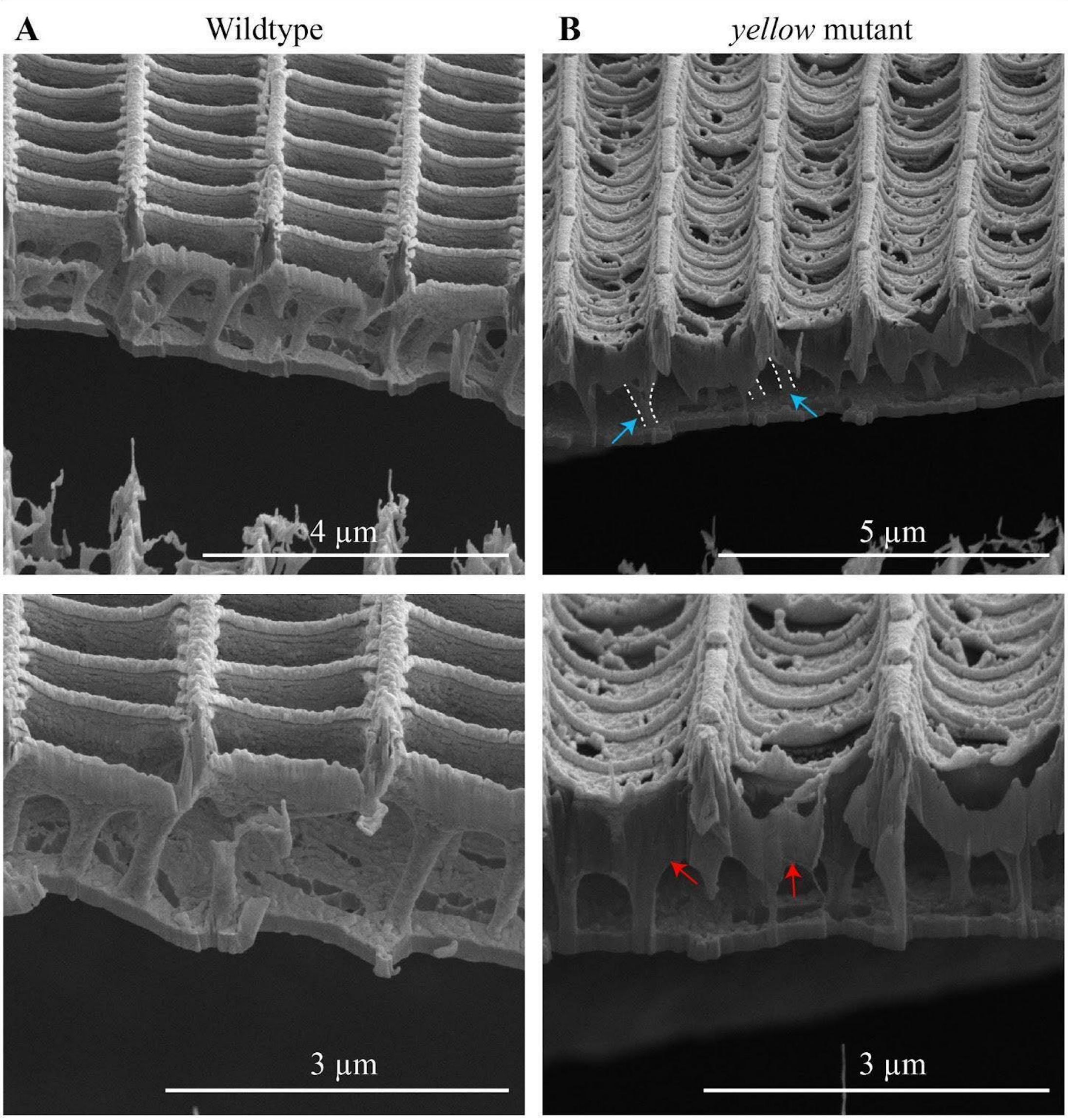

**Fig. S6. FIB-SEM cross-sections of *Bicyclus anynana* wildtype and *yellow* mutant scales.** (A) In the wildtype cross-section, the trabeculae are straight, and the extensions of the cross-ribs are less prominent, which is consistent with the highly pigmented *Junonia orithya* female brown scales. (B) In the *yellow* mutant cross-sections (top), blue arrows and dotted white lines indicate curved trabeculae, and (bottom) red arrows indicate sheet-like extensions of the cross-ribs. The less pigmented *yellow* mutant scales show sheet-like extensions of the cross-ribs, similar to those of less pigmented *Junonia orithya* male blue scales.

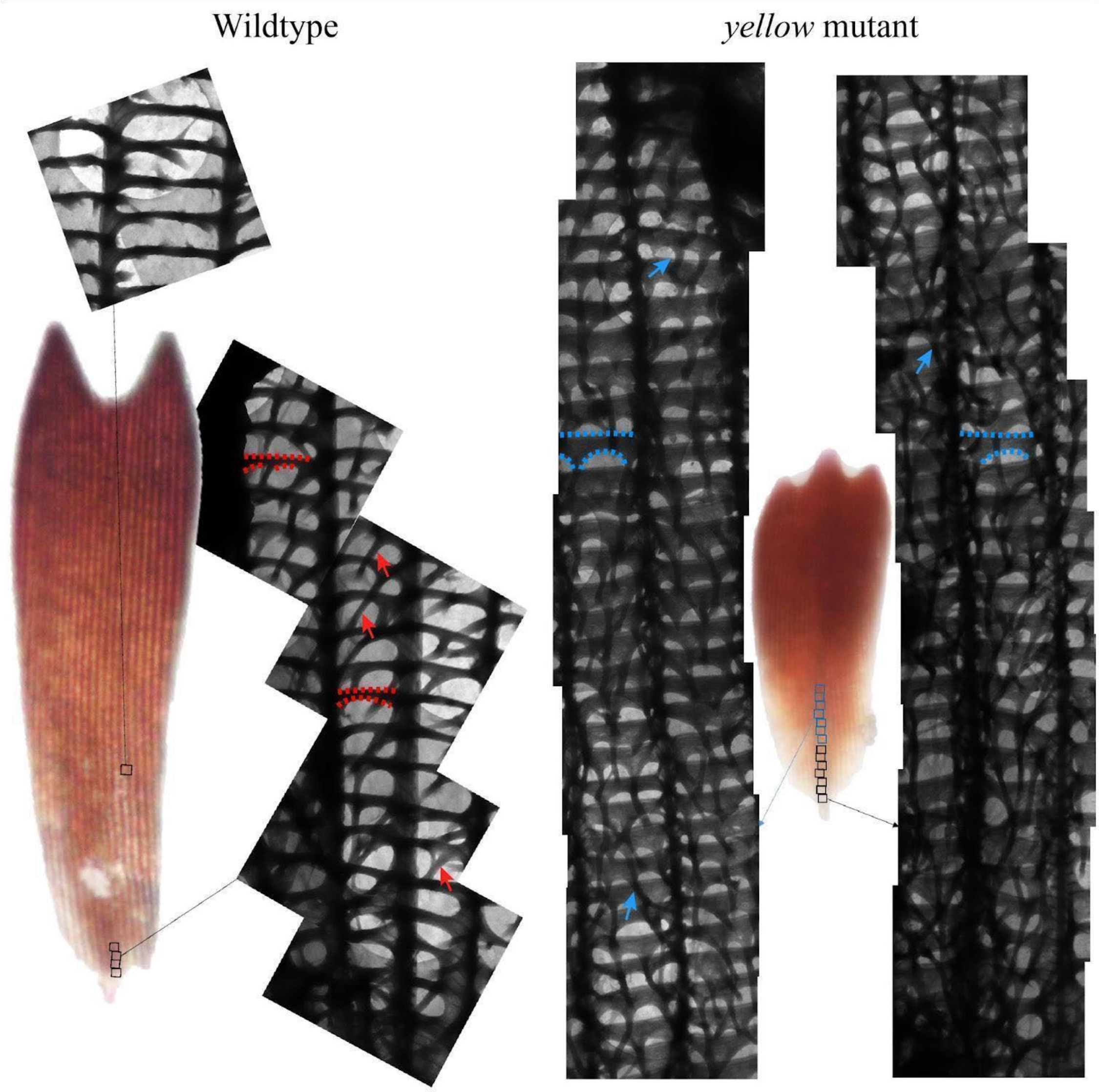

**Fig. S7. TEM top-down images of *Bicyclus anynana* wildtype and *yellow* mutant scales.** Entire scales were directly mounted on holey carbon TEM grids and imaged. Images were taken in series at the different locations indicated by square boxes. Wildtype scales have thin cross-ribs (red dotted lines) with straight trabeculae (red arrows) extending to the lower lamina. In comparison to the wildtype scales, *yellow* mutant scales have sheet-like extensions of the cross-ribs (blue dotted lines) and curved trabeculae (blue arrows).

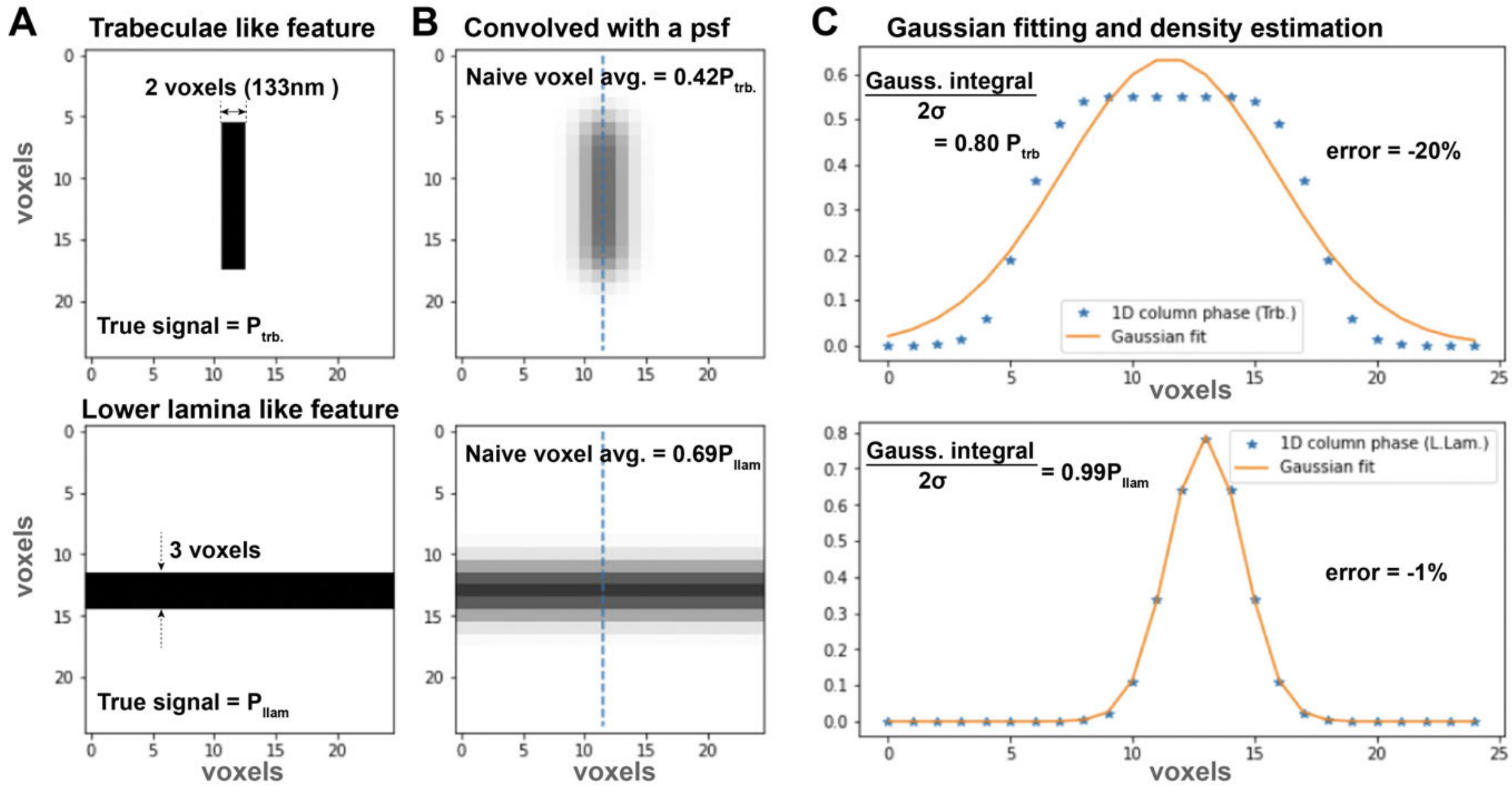

**Fig. S8. Error bar analysis between a vertically thin structure and a horizontally thin structure matching minimum dimensions of *Junonia* male blue scale trabeculae and lower lamina.** (A) The cuboidal structures represent trabeculae and lower lamina (voxel size=66.5nm). (B) The data blurred with a Gaussian PSF (sigma=1.25 voxels). (C) The Gaussian fitting of the blue dotted profile from panel B. The minimum dimensions of the upper lamina features and the lower lamina are comparable across all scales (~150nm) except for the *Junonia* male blue scale. In this particular case, the upper lamina features are approximately two voxels wide laterally (140-150 nm), while the lower lamina thickness is 190 nm, which is almost 1 voxel more. To assess the potential impact of this extreme case, we simulated cuboidal features with these dimensions and convolved them with a PSF to mimic the PXCT voxels. This analysis does not explicitly account for other potential confounders such as noise and reconstruction artifacts, making it a conservative estimate of error under the assumption that all such effects are encapsulated within a single effective PSF. Our results show that if voxel intensities are naively averaged, only ~42% of the true signal is recovered for trabeculae-like features and ~69% for lower lamina-like features. However, our Gaussian fitting approach applied to 1D voxel columns significantly improves these estimates, which recovers ~80% and ~99% of the true signal, respectively. This corresponds to error bars of 20% and 1%. While there remains a 19% difference in estimation error between these two specific feature types, this discrepancy is insufficient to undermine our conclusions. Our lower-bound mean density estimate for the trabeculae is 0.88 $g/cm^3$ and 1.24 $g/cm^3$ for the lower lamina in the *Junonia* male blue scale. Specifically, the observed ~30% density difference between the lower lamina and the densest upper lamina feature trabeculae in the *Junonia* male blue scale remains valid even when this modeling uncertainty is considered. In all other scales analyzed, the lower laminae thicknesses are 150-160 nm, and the upper lamina features are similarly dimension; therefore, the error bars are similar across the features, which do not affect the interpretation of relative density trends.

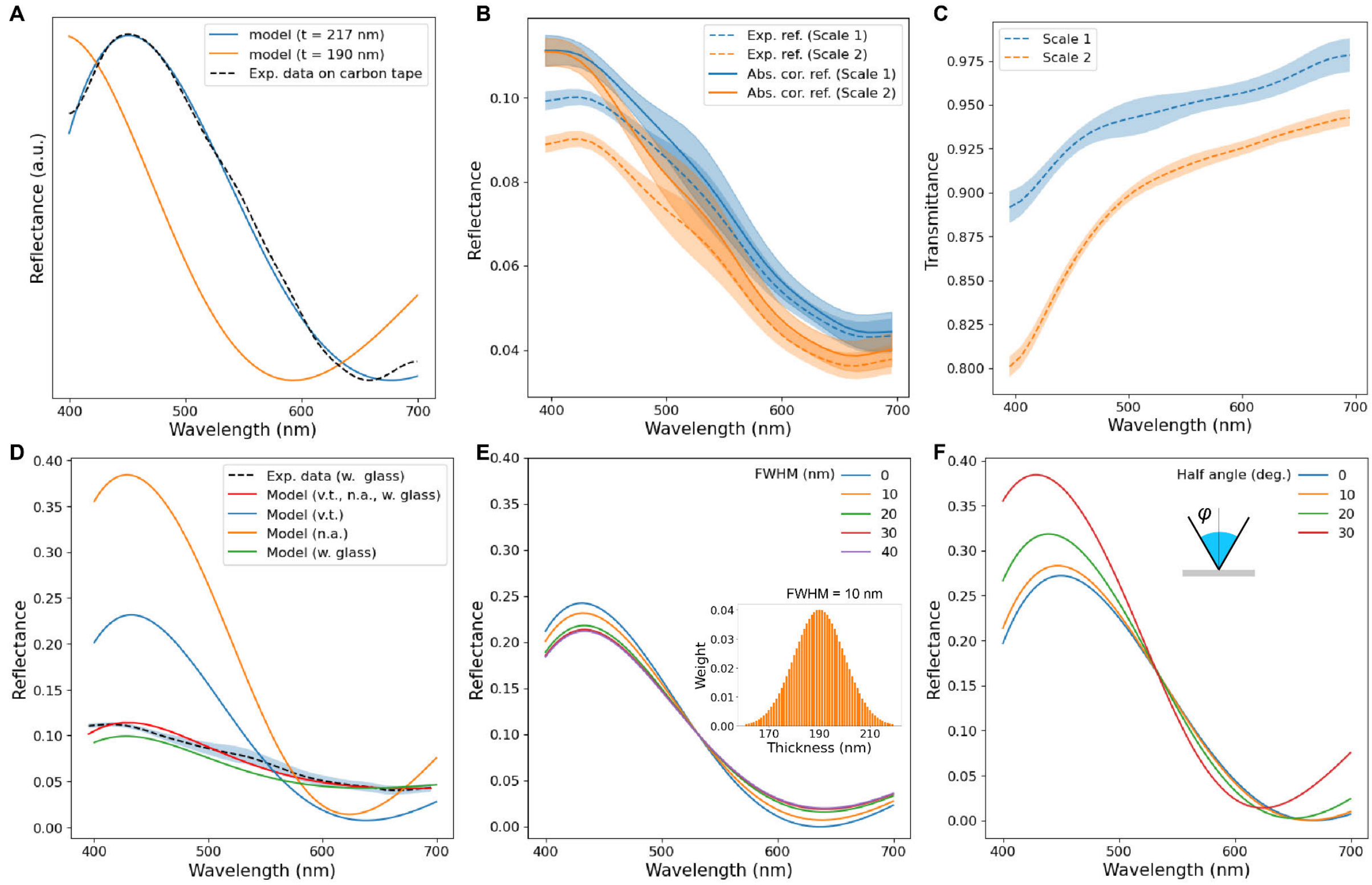

**Fig. S9. *Junonia* male blue scale reflectance and transmittance.** (A) The thin-film interference model shows that a single-layered material with a wavelength-independent refractive index of 1.56 must be 217 nm thick to match the peak and trough of an experimental reflectance measurement of the scale on a black carbon tape. (B) The experimental reflectance measurements from two different scales (same individual and on same measurement conditions) on a glass slide and their corresponding absorbance corrected reflectance spectra. The difference in the experimental reflectance measurements between the scales is partially due to the absorbance; thus, the difference between the absorbance-corrected spectra is smaller. (C) The transmittance measurements of the respective scales were determined from their absorbance measurements. The relative difference (~10%) in the reflectance between the scales corresponds well with their relative difference in the transmittance. This validates the need to account for the absorbance in the thin-film interference model. (D) The absorbance-corrected experimental reflectance data measured on a glass slide is compared with a model that accounts for the variation in thickness (v.t.), a model that accounts for the numerical aperture (n.a.), a model that includes the glass slide (w. glass) and a model that accounts for all three cases. (E) Reflectance spectra calculated for a thin film with an average thickness of 190 nm and thickness varying in a Gaussian way (as shown in the inset), with full-width at half maximum (FWHM) 0, 10, 20, 30, and 40 nm. Though the numerical aperture and variation in the thickness cause the troughs to have an offset from zero, the backscattering from the glass slide provides a substantial offset. (F) Reflectance spectra calculated for a thin film with a thickness of 190 nm and different numerical apertures with a maximum half-angle ($\varphi$) of 0, 10, 20, and 30 (degrees). Since variations in thickness, numerical aperture, and the glass slide beneath all influence the reflectance spectra, we included their effects by using the measured values while fitting for the refractive indices of the multi-layered lower lamina.

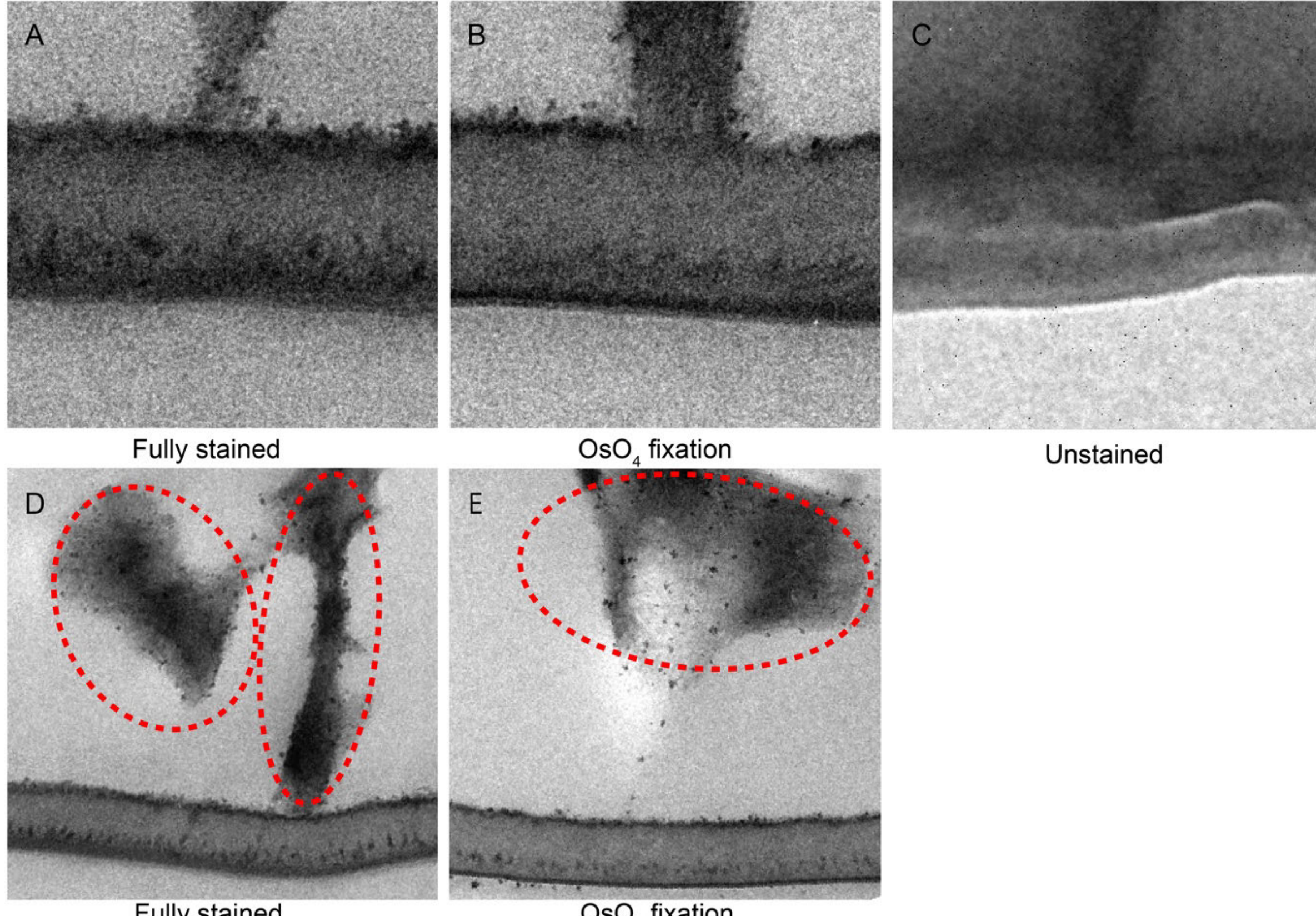

**Fig. S10. TEM resin sections.** (A, D) Postfixation with osmium tetroxide followed by secondary staining with lead citrate and uranyl acetate. (B, E) Only a postfixation with osmium tetroxide but no secondary stains. (C) Completely unstained. In both (D) and (E), with and without lead staining, contrast is consistently strong in the upper lamina suggesting even penetration of the heavy metal stain throughout the scale structure. We found that the secondary staining shows the multi-layered lower lamina but has a lot of darker patches caused by the stains. The image without secondary staining clearly distinguishes between the layers, due to the biased attachment of osmium tetroxide when used as post-fixative. Osmium tetroxide's preference for lipids and melanin shows a distinct contrast between the melanin-rich, chitin-protein layers, and lipid-rich layers [3,4].

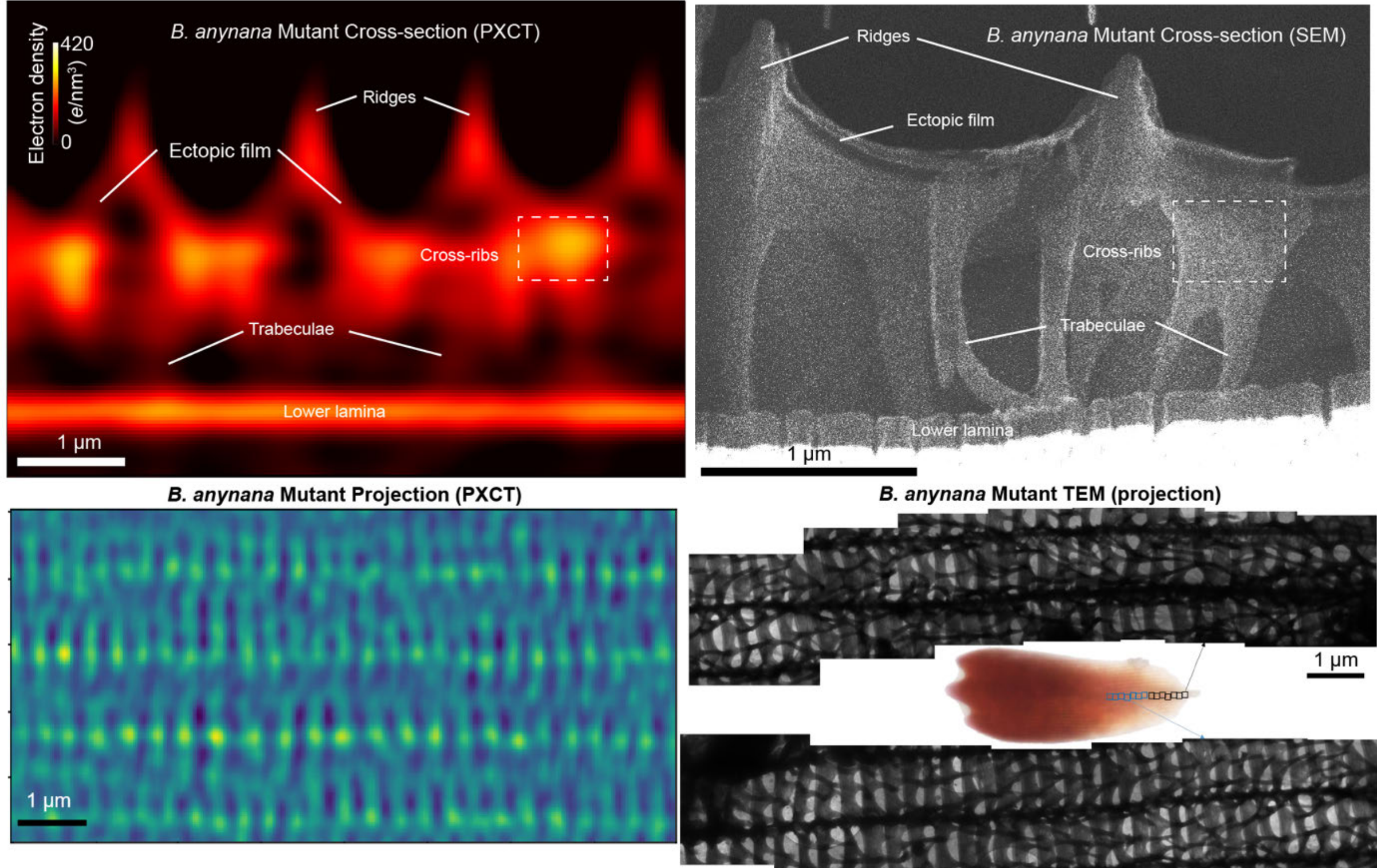

**Fig. S11. Complex architecture of the *Bicyclus anynana yellow* mutant scale.** PXCT provides a medium-resolution 3D density volume that can be directly correlated with high-resolution electron micrographs to understand the complex architecture of the *yellow* mutant scale. PXCT cross-section (top left) is compared with the SEM cross-section (top right), revealing sheet-like cross-ribs (dashed box) spanning between the ridges and the curved trabeculae. A line-integral PXCT projection (bottom left) is compared with stitched top-down TEM images (bottom right), confirming that these cross-ribs are tilted and densely packed.

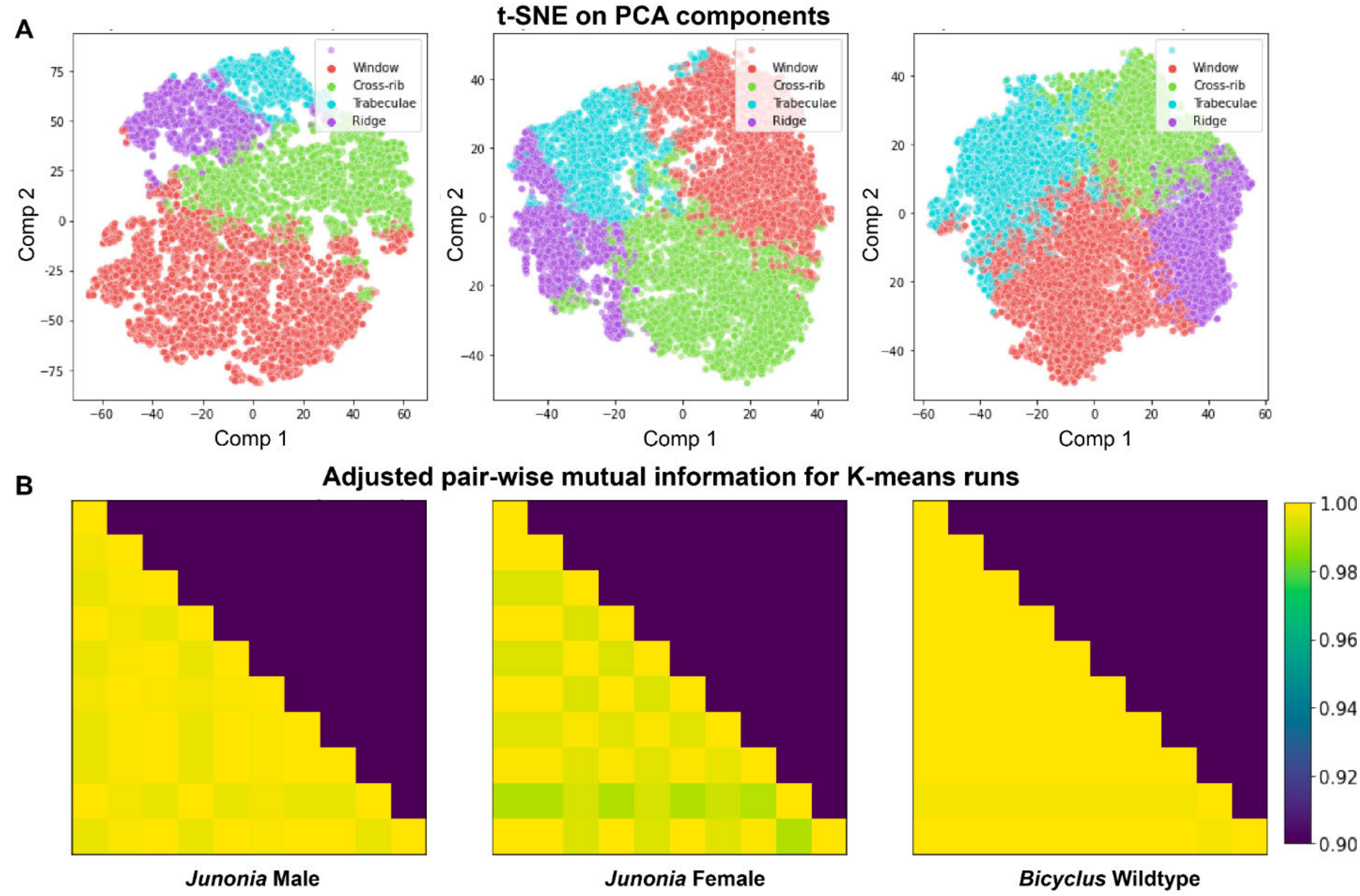

**Fig. S12. Validation for the K-means clustering of sub-structures within the upper lamina.** (A) A t-SNE dimensionality reduction is carried out on the PCA components, and the two-dimensional output is plotted with the corresponding labels from the K-means clustering. Though the K-means clustering results are grouped together in the t-SNE output, the clusters are not well separated. (B) We ran 10 K-means clustering with the different random states for the initialisation on each scale and computed the adjusted pairwise mutual information metric. The high mutual information across all the runs indicates that the K-means clusterings are consistent and repeatable.